\documentclass[prd,aps,nofootinbib,superscriptaddress]{revtex4-2}
\usepackage{epsfig}
\usepackage{amsmath, slashed}
\usepackage{booktabs}

\usepackage{array}
\usepackage{appendix}
\usepackage{braket}
\usepackage{amssymb}
\usepackage{dutchcal}
\usepackage{ifsym}
\usepackage{comment}
\usepackage{tablefootnote}
\usepackage{longtable} 
\usepackage{mathtools}
\usepackage{amssymb}

\usepackage[table,xcdraw]{xcolor}

\usepackage{hhline}
\usepackage[most]{tcolorbox}
\usepackage{float}

\usepackage[normalem]{ulem}

\renewcommand{\arraystretch}{1.5}
\usepackage{bm}
\usepackage{amsfonts}
\definecolor{myGreen}{rgb}{0.2,0.72,0.2}
\definecolor{tableGray}{rgb}{0.9,0.9,0.9}
\definecolor{tableblue}{rgb}{0.9,0.9,1.0}

\renewcommand{\[}{\begin{equation}}
\renewcommand{\]}{\end{equation}}

\newcommand{\ta}{\left(}
\newcommand{\tc}{\right)}

\definecolor{pine}{rgb}{0.0, 0.5, 0.0}

\allowdisplaybreaks

\makeatletter

    \def\CT@@do@color{%
      \global\let\CT@do@color\relax
            \@tempdima\wd\z@
            \advance\@tempdima\@tempdimb
            \advance\@tempdima\@tempdimc
    \advance\@tempdimb\tabcolsep
    \advance\@tempdimc\tabcolsep
    \advance\@tempdima2\tabcolsep
            \kern-\@tempdimb
            \leaders\vrule
                    \hskip\@tempdima\@plus  1fill
            \kern-\@tempdimc
            \hskip-\wd\z@ \@plus -1fill }
    \makeatother
\begin{document}

\title{The valence quark, sea,  and gluon content of the pion from the parton distribution functions and the electromagnetic form factor }
  
  \author{Barbara Pasquini}
\thanks{Electronic address: barbara.pasquini@unipv.it }
\affiliation{Dipartimento di Fisica, Universit\`a degli Studi di Pavia, I-27100 Pavia, Italy}
\affiliation{Istituto Nazionale di Fisica Nucleare, Sezione di Pavia, I-27100 Pavia, Italy}

\author{Simone Rodini}
\thanks{Electronic address: simone.rodini@polytechnique.edu}
\affiliation{CPHT, CNRS, Ecole Polytechnique, Institut Polytechnique de Paris, Route de Saclay, 91128 Palaiseau, France}

\author{Simone Venturini}
\thanks{Electronic address: simone.venturini01@universitadipavia.it }
\affiliation{Dipartimento di Fisica, Universit\`a degli Studi di Pavia, I-27100 Pavia, Italy}
\affiliation{Istituto Nazionale di Fisica Nucleare, Sezione di Pavia, I-27100 Pavia, Italy}


\date{\today}

\collaboration{
MAP Collaboration}\thanks{The MAP acronym stands for ``Multi-dimensional Analyses of Partonic distributions''. It refers to a collaboration aimed at studying the three-dimensional structure of hadrons. }

\begin{abstract}
We present a  light-front model calculation 
of the pion parton distribution functions (PDFs) and the pion electromagnetic form factor. The pion state is modeled in terms of light-front wave functions (LFWFs) for 
the  $q\bar q$,  $q\bar q q\bar q$,  $q\bar q g$, and $q\bar q gg$ components. 
We design the LFWFs so that the parameters in the longitudinal and transverse momentum space enter separately in the fit of the pion PDFs and the electromagnetic form factor, respectively. 
We extract the pion PDFs  within the xFitter framework using  available Drell-Yan and 
photon-production data. With the obtained parameters in the longitudinal-momentum space, we then fit the available experimental  data on the pion electromagnetic form factor to constrain the remaining parameters in the transverse-momentum space. 
The results for the pion PDFs are compatible with existing extractions and lattice calculations, and the fit to the pion electromagnetic form factor data works quite successfully.
The obtained parametrization for the LFWFs marks a step forward towards a unified description of different  hadron distribution functions in both the longitudinal- and transverse-momentum space and will be further applied to a phenomenological study of transverse-momentum dependent parton distribution functions and generalized parton distributions.
\end{abstract}

\maketitle

\section{Introduction}
\label{Sec_Introduction}

A successful approach in high-energy scattering
is based on light-front quantization where hadrons are described by light-front wave functions (LFWFs)~\cite{Brodsky:1997de}. The
latter are expressed as an expansion of various quark $(q)$, antiquark $(\bar q)$, and gluon ($g$) Fock components. Schematically,
a pion state is conceived as the following superposition
\begin{eqnarray}
|\pi\rangle=\psi_{q\bar q}\,|q\bar q\rangle +\psi_{q\bar q q\bar q}\, | q\bar qq\bar q\rangle+\psi_{q\bar q g}\,|qqg\rangle + \psi_{q\bar q gg}\, |q\bar q gg\rangle+\dots\, ,
\end{eqnarray}
where, in the light-cone gauge $A^+=0$, the  LFWF $\psi$ for each parton configuration involves a number of independent amplitudes
corresponding to different combinations of quark orbital angular momentum and helicity~\cite{Ji:2003yj}.
The light-front representation has a number of simplifying properties. 
In particular, it allows one to describe the hadronic matrix elements which parametrize
the soft contribution in inclusive and exclusive reactions in terms of overlap of LFWFs
with different parton configurations~\cite{Lorce:2011dv}. 
A variety of phenomenological and theoretical models have been devised to give explicit expressions for the
LFWF amplitudes 
and to access information on the internal structure of the pion from different partonic functions.
In this work, we will restrict ourselves to investigating  the pion parton distributions (PDFs) and pion electromagnetic (e.m.) form factor.

Most  models for the pion PDFs confined the study to the  valence-quark distributions or included  a dynamical gluon at the hadronic scale of the model with the sea quark contribution  generated perturbatively by QCD evolution~\cite{deMelo:2005cy,deMelo:2008rj,Frederico:2009fk,Chang:2013pq,Pasquini:2014ppa,Gutsche:2014zua,Lorce:2016ugb,Chouika:2017rzs,Bacchetta:2017vzh,deTeramond:2018ecg,Watanabe:2019zny,Lan:2019vui,Qian:2020utg,Bastami:2020asv,Raya:2021zrz,Chavez:2021llq,Li:2021cwv,Lan:2021wok,Ydrefors:2021dwa,dePaula:2022pcb,Li:2022mlg,Lu:2022cjx,Cui:2022bxn,Cui:2021mom,Zhu:2023lst}.
In the last few years, 
there has been also an increasing number of calculations of $x$-dependent pion PDF in lattice QCD following different approaches. They have been mainly restricted to the valence-quark sector~\cite{Gao:2022iex,Karthik:2021qwz,Lin:2020ssv,Gao:2020ito,Sufian:2020vzb,Sufian:2019bol,Izubuchi:2019lyk,Joo:2019bzr,Izubuchi:2019lyk,Zhang:2018nsy} and only recently have been extended to the gluon sector~\cite{Fan:2021bcr}.
Complementing these theory developments, global analyses of pion PDFs have been performed mostly using pion-induced Drell-Yan (DY) data and $J/\psi$-production data or direct photon production~\cite{Owens:1984zj,Aurenche:1989sx,Sutton:1991ay,Gluck:1991ey,Gluck:1999xe,Bourrely:2022mjf,Bourrely:2020izp,Bourrely:2018yck,Chang:2020rdy,Barry:2018ort,Barry:2021osv,JeffersonLabAngularMomentumJAM:2022aix}. However, the current knowledge of the pion PDFs is less accurate than for the nucleon PDFs, mainly because there are fewer experimental data to constrain the pion PDFs, especially for the sea-quark and gluon contributions.
New experiments are expected to further our knowledge of the pion PDFs.
The planned experiments  at Jefferson Lab~\cite{Arrington:2021alx} and at new-generation facilities using high-luminosity electron-proton collisions~\cite{AbdulKhalek:2021gbh,Anderle:2021wcy} are developing  the capacity to access the pion PDFs via the Sullivan process~\cite{Sullivan:1971kd}, consisting of scattering off the pion in proton to pion fluctuations~\cite{Aguilar:2019teb,Arrington:2021biu}. Furthermore, experiments from COMPASS++/AMBER will exploit  high-energy, high-intensity pion  beams to probe directly the partonic structure of the pion~\cite{Adams:2018pwt}.
 
 Different insights into the pion structure can be gained from the study of the pion e.m. form factor.
 The e.m. form factor probes the charge distribution in the pion and is  a good observable for studying the onset, with increasing
energy, of the perturbative QCD regime for exclusive processes~\cite{Lepage:1979zb,Efremov:1979qk}.
It  has been successfully described in a variety of light-front quark models~\cite{Lan:2021wok,Ydrefors:2021dwa,Gutsche:2014zua,Frederico:2009fk,deMelo:2005cy,deMelo:2003uk,deMelo:2002yq,Hwang:2001hj,Bakker:2000pk,Cardarelli:1995hn,Cardarelli:1995dc,Frederico:1992ye,Chung:1988mu} and  has witnessed enormous progress in recent lattice QCD calculations~\cite{Gao:2021xsm,Alexandrou:2021ztx,Wang:2020nbf,ETM:2017wqc,Aoki:2015pba,Koponen:2015tkr,Fukaya:2014jka,Brandt:2013dua,Nguyen:2011ek,JLQCD:2009ofg,Frezzotti:2008dr,Boyle:2008yd,QCDSFUKQCD:2006gmg,Bonnet:2004fr,Feng:2019geu}.
Data for the pion e.m. form factor at low momentum transfer ($Q^2\leq 0.253$ GeV$^2$) have been measured at Fermilab~\cite{Dally:1981ur,Dally:1982zk} and CERN~\cite{Amendolia:1984nz,NA7:1986vav}  by scattering of pions off atomic electrons and were used to extract  also the pion charge radius. At higher momentum transfer, up to $Q^2\approx 10$ GeV$^2$, the pion e.m. form factor was extracted in experiments of pion electroproduction in Cornell~\cite{Bebek:1974ww,Bebek:1976qm,Bebek:1977pe}, DESY~\cite{Brauel:1979zk,Ackermann:1977rp}, and JLab~\cite{JeffersonLabFpi:2000nlc,JeffersonLabFpi:2007vir,JeffersonLabFpi-2:2006ysh,JeffersonLab:2008gyl,JeffersonLab:2008jve}, by exploiting the Sullivan mechanism. New accurate data are expected at intermediate values of $Q^2$ from upcoming JLab measurements~\cite{Arrington:2021alx} while the future electron-ion colliders will potentially give access to the region at higher  momentum transfer, up to $Q^2=30$ GeV$^2$.
 
In this work,
we propose a new parametrization of the pion LFWFs which comprises the Fock states of the $q\bar q$,  $q\bar q q\bar q$,  $q\bar q g$, and $q\bar q gg$ components and is adapted to reproduce simultaneously the available experimental data on the pion PDFs and e.m. form factor.
To our knowledge, the expansion in the Fock space up to the two-gluon component  represents the largest basis that has been used so far in  light-front model calculations of the pion PDFs and e.m. form factor.
Furthermore, we design the LFWFs so that the parameters in the longitudinal and transverse momentum space enter separately in the fit of the pion PDFs and e.m. form factor, respectively. 
In particular, the parametrization in the longitudinal momentum space is dictated by the pion distributions amplitudes, while the functional form in the transverse-momentum space is constrained so that the
LFWF overlap representation of 
the pion PDFs  does not depend on the transverse-momentum dependent parameters. 
The fit to the  experimental data of the pion PDF is performed  using the open-source tool xFitter~\cite{Alekhin:2014irh} which was recently extended to extract the pion PDF~\cite{Novikov:2020snp}. With the obtained parameters in the longitudinal-momentum space, we then fit the available experimental  data on the pion e.m. form factor to constrain the  parameters in the transverse-momentum space. 
Our results for the valence, sea, and gluon contribution to the pion PDFs are  
consistent with recent extractions~\cite{Novikov:2020snp,Barry:2018ort,Bourrely:2022mjf}, although the considered set of experimental data does not constraint well the sea and gluon contributions.
Furthermore, the fit to the available experimental data for the pion e.m. form factor works very successfully,  proving the merit of the adopted strategy to build the parametrization of the LFWFs.  

The paper is organized as follows. After a brief review  of the light-front Fock-space expansion of the pion state in Sec.~\ref{Sec_LFWAs}, we construct  the explicit parametrization for the LFWFs  in Sec.~\ref{Sec_Model}.
The pion PDFs are discussed in Sec.~\ref{Sec_PDFs},  where we present the model expressions  of the pion PDFs obtained through  the LFWF overlap representation. We then summarize the fit procedure of the pion PDFs within the xFitter framework and discuss the results in comparison with other recent extractions and model calculations.
Sec.~\ref{Section_FF} is dedicated to 
 the pion e.m. form factor, discussing  the results from the fit to extant experimental data.
In Sec.~\ref{Sec_Conclusions} we summarize our results and give an
outlook. Technical details about the
LFWF overlap representation of the pion PDF and e.m. form factor are given in App.~\ref{App_LFWA_PDFs} and App.~\ref{App_LFWA_FFs}, respectively.

\section{Light Front Wave Amplitudes}\label{Sec_LFWAs}
In this section, we review the classification of the light-front wave functions of the pion, considering the Fock-space configuration up to four partons, i.e.
\begin{equation}
 \label{Eq_Split_Pion_State} \ket{\pi(P)} = \ket{\pi(P)}_{q \bar{q}} + \ket{\pi(P)}_{q \bar{q}g} + \ket{\pi(P)}_{q \bar{q}gg} + \sum_{\left\{ \mathcal{s} \bar{\mathcal{s}}\right\}}\ket{\pi(P)}_{q \bar{q} \left\{ \mathcal{s} \bar{\mathcal{s}}\right\}},
\end{equation}
where $q=u,d$ and  the sum in $\left\{\bar{\mathcal{s}} \mathcal{s}\right\}$ runs over the $N_f$-flavor pairs of the sea quarks ($u\bar{u},$ $d\bar{d},$ $s\bar{s}$ \sout{in this work} at the model scale).
The LFWF for each parton configuration can be classified  according to the total parton light-cone helicity $\lambda$ or, equivalently, to the angular
momentum projection $l_z=-\lambda$, which follows from
angular momentum conservation~\cite{Ji:2003yj}. In principle, the states up to  four partons in Eq.~\eqref{Eq_Split_Pion_State} involve 94 independent light-front wave amplitudes (LFWAs), corresponding to all the possible
combinations of parton helicities. In order to keep the model as simple as possible,
in our analysis  we restrict ourselves  to consider only the projection on the $l_z=0$ component, i.e.
\begin{align}
\label{Eq_Split_Pion_State_Lz0}
    \ket{\pi(P)}^{l_z = 0} = \ket{\pi(P)}_{q \bar{q}}^{l_z = 0} + \ket{\pi(P)}_{q \bar{q}g}^{l_z = 0} + \ket{\pi(P)}_{q \bar{q}gg}^{l_z = 0} +\sum_{\left\{ \mathcal{s} \bar{\mathcal{s}}\right\}}\ket{\pi(P)}_{q \bar{q} \left\{ \mathcal{s} \bar{\mathcal{s}}\right\}}^{l_z = 0}.
\end{align}

Compared to the original classification in Ref.~\cite{Ji:2003yj}, we make the further simplification of neglecting the  LFWAs that are multiplied by  coefficients depending on the parton transverse momenta,  in order to have simpler relations to the pion distribution amplitudes, as discussed in Sec.~\ref{Sec_Model}.
We then have the following expressions:

\begin{align}
    \label{Eq_qq_State_lz=0}
    \ket{\pi(P)}_{q \bar{q}}^{l_z = 0} &=  \int \text{d[1]}\text{d[2]}  \frac{\delta_{c_{1} c_{2}}}{\sqrt{3}}\psi^{(1)}_{q \bar{q}}(1,2) \left[q^{\dagger}_{c_{1} \uparrow}(1) \bar{q}^{\dagger}_{c_{2} \downarrow}(2) - q^{\dagger}_{c_{1} \downarrow}(1) \bar{q}^{\dagger}_{c_{2} \uparrow}(2) \right] \ket{0},
\\
    \label{Eq_qqg_State}
    \ket{\pi(P)}_{q \bar{q} g}^{l_z = 0} &= \int \text{d[1]}\text{d[2]}\text{d[3]}  \frac{T^{a}_{c_1 c_2}}{2}\psi^{(1)}_{q \bar{q} g}(1,2,3) \left[ \left(q \bar{q}\right)^{\dagger}_{A, 1} g_{a \downarrow}^{\dagger}\left(3\right) - \left(q \bar{q}\right)^{\dagger}_{A, -1} g_{a \uparrow}^{\dagger}\left(3\right) \right]\ket{0},\\
    \label{Eq_qqgg_State}
    \nonumber \ket{\pi(P)}_{q \bar{q} gg}^{l_z = 0} &=  \int \text{d[1]}\text{d[2]}  \text{d[3]} \text{d[4]}  \frac{\delta_{c_1 c_2}\delta^{a b}}{\sqrt{24}} \biggl\{\psi^{(1)}_{q \bar{q} gg}(1,2,3,4) \left(q \bar{q}\right)^{\dagger}_{A, 0} \left(gg\right)^{\dagger}_{S,0} \\
    & + \psi^{(2)}_{q \bar{q} gg}\left(1,2,3,4\right)\left(q \bar{q}\right)^{\dagger}_{S,0}\left(gg\right)^{\dagger}_{A,0}\biggr\} \ket{0},
\\
    \label{Eq_qqqq_State}
    \nonumber \ket{\pi(P)}_{q \bar{q} \left\{\mathcal{s}\bar{\mathcal{s}}\right\}}^{l_z = 0} &=  \int \text{d[1]}\text{d[2]}\text{d[3]}\text{d[4]}  \frac{\delta_{c_1 c_2}\delta_{c_3 c_4}}{3} \biggl\{\psi^{(1)}_{q \bar{q} \mathcal{s} \bar{\mathcal{s}}}(1,2,3,4) \left(q \bar{q}\right)^{\dagger}_{A, 0} \left(\mathcal{s} \bar{\mathcal{s}}\right)^{\dagger}_{S,0} \\
    \nonumber & + \psi^{(2)}_{q \bar{q} \mathcal{s} \bar{\mathcal{s}}}\left(1,2,3,4\right)\left(q \bar{q}\right)^{\dagger}_{S,0}\left(\mathcal{s} \bar{\mathcal{s}}\right)^{\dagger}_{A,0}\\
    &+ \psi^{(3)}_{q \bar{q} \mathcal{s} \bar{\mathcal{s}}}\left(1,2,3,4\right)\left[\left(q \bar{q}\right)^{\dagger}_{A,1}\left(\mathcal{s} \bar{\mathcal{s}}\right)^{\dagger}_{A,-1} - \left(q \bar{q}\right)^{\dagger}_{A,-1}\left(\mathcal{s} \bar{\mathcal{s}}\right)^{\dagger}_{A,1}\right] \biggr\} \ket{0},
\end{align}
where 
$q^\dagger_{c\lambda}$ and $\bar{q}^\dagger_{c\lambda}$ are creation operators of a quark and antiquark with flavor $q$, helicity $\lambda$ and color $c$,
respectively. The LFWAs $\psi$  are functions of parton momenta with arguments $i=(x_i,\boldsymbol{k}_{\perp i})$ representing  the fraction of longitudinal parton momentum $x_i=k^+_i/P^+$ and the transverse parton momentum $\boldsymbol{k}_{\perp i}$.
Furthermore, in Eqs.~\eqref{Eq_qq_State_lz=0}-\eqref{Eq_qqqq_State}, $T^a_{i j} = \frac{\lambda^a_{i j}}{2}$ are the $SU(3)$ color matrices  and the following operators have been introduced: \begin{align}
    \label{Eq_Operatori_Simmetrici/Asimmetrici}
    \left(q \bar{q}\right)^{\dagger}_{S,0} & = q^{\dagger}_{c_{1} \uparrow}(1)\bar{q}^{\dagger}_{c_2 \downarrow}(2) + q^{\dagger}_{c_{1} \downarrow}(1)\bar{q}^{\dagger}_{c_2 \uparrow}(2), \qquad  \left(\mathcal{s} \bar{\mathcal{s}}\right)^{\dagger}_{S,0} = \mathcal{s}^{\dagger}_{c_{3} \uparrow}(3)\bar{\mathcal{s}}^{\dagger}_{c_4 \downarrow}(4) + \mathcal{s}^{\dagger}_{c_3 \downarrow}(3)\bar{\mathcal{s}}^{\dagger}_{c_4 \uparrow}(4), \\
    \left(q \bar{q}\right)^{\dagger}_{A,0} & = q^{\dagger}_{c_{1} \uparrow}(1)\bar{q}^{\dagger}_{c_{2} \downarrow}(2) - q^{\dagger}_{c_{1} \downarrow}(1)\bar{q}^{\dagger}_{c_2 \uparrow}(2), \qquad \left(\mathcal{s} \bar{\mathcal{s}}\right)^{\dagger}_{A,0} = \mathcal{s}^{\dagger}_{c_3 \uparrow}(3)\bar{\mathcal{s}}^{\dagger}_{c_4 \downarrow}(4) - \mathcal{s}^{\dagger}_{c_3 \downarrow}(3)\bar{\mathcal{s}}^{\dagger}_{c_4 \uparrow}(4),\\
    \left(q \bar{q}\right)^{\dagger}_{A,1} & = q^{\dagger}_{\uparrow c_1}(1) \bar{q}^{\dagger}_{c_2 \uparrow}(2), \qquad \qquad \qquad \qquad \quad \left(\mathcal{s} \bar{\mathcal{s}}\right)^{\dagger}_{A,1} = \mathcal{s}^{\dagger}_{\uparrow c_3}(3) \bar{\mathcal{s}}^{\dagger}_{c_4 \uparrow}(4),\\
    \left(q \bar{q}\right)^{\dagger}_{A,-1} & = q^{\dagger}_{\downarrow c_1}(1) \bar{q}^{\dagger}_{c_2 \downarrow}(2), \qquad \qquad \qquad \qquad \quad \left(\mathcal{s} \bar{\mathcal{s}}\right)^{\dagger}_{A,-1} = \mathcal{s}^{\dagger}_{\downarrow c_3}(3) \bar{\mathcal{s}}^{\dagger}_{c_4 \downarrow}(4),\\
    \left(g g\right)^{\dagger}_{S,0} & = g^{\dagger}_{a \uparrow}(3) g^{\dagger}_{b \downarrow}(4)+ g^{\dagger}_{a \downarrow}(3) g^{\dagger}_{b \uparrow}(4), \\
    \left(g g\right)^{\dagger}_{A,0} & = g^{\dagger}_{a \uparrow}(3) g^{\dagger}_{b \downarrow}(4)- g^{\dagger}_{a \downarrow}(3) g^{\dagger}_{b \uparrow}(4) .
\end{align}

The integration measure in Eqs.~\eqref{Eq_qq_State_lz=0}-\eqref{Eq_qqqq_State} is defined as
\begin{equation}
\label{Eq_Measures}
\displaystyle \prod_{i=1}^N \text{d[}i\text{]} = \left[dx\right]_N \left[d^2 \boldsymbol{k}_{\perp}\right]_N,
\end{equation}
where
\begin{equation}
    \label{Eq_Measure_xN}
    \left[dx\right]_N = \prod_{i=1}^{N}\frac{ dx_i}{  \sqrt{ x_i}}  \delta\left(1 - \displaystyle\sum_{i=1}^{N} x_i\right),
\quad
   \quad
    \left[d^2 \boldsymbol{k}_{\perp} \right]_{N} = \frac{1}{[2(2 \pi)^3]^{N-1}}\prod_{i=1}^{N}d^2 \boldsymbol{k}_{\perp i} \delta^{(2)}\left(\sum_{i=1}^{N}\boldsymbol{k}_{\perp i}\right).
\end{equation}

\section{Model for the LFWAs} \label{Sec_Model}
In order to construct a model with a realistic structure for the LFWAs, we will exploit their connection to the pion distribution amplitudes (DAs), which are pion-to-vacuum transition matrix elements of collinear operators~\cite{Radyushkin:1977gp,Chernyak:1977as,Chernyak:1980dj,Chernyak:1977fk,Chernyak:1980dk}. 
Without loss of generality, we can write any LFWA as:
\begin{equation}
    \label{Eq_Factorized_LFWA}
    \psi^{(i)}\left(1, 2, \dots , N\right) = \phi^{(i)}\left(x_1, x_2, \dots , x_N\right) \Omega^{(i)}_{N,\beta}\left(x_1, \boldsymbol{k}_{\perp 1}, x_2, \boldsymbol{k}_{\perp 2}, \dots,  x_N, \boldsymbol{k}_{\perp N} \right),
\end{equation}
where we introduced the label $\beta$ for the parton composition $\{q\bar q,q\bar q g,  q\bar q \mathcal{s}\bar{\mathcal{s}}, q\bar q gg\}$.

The integral over the intrinsic transverse momenta of the LFWA for $l_z=0$ state with $N$ partons can be expressed as a linear combination of DAs of $N$ partons of matching type. Schematically 
\begin{equation}
\int \left[ d^2 \boldsymbol{k}_{\perp} \right]_N \psi^{(i)}\left(1, 2, \dots , N\right)  = \sum_{j} a_{ij} d_j(x_1,...,x_N),
\label{Fundamental_LFWA_to_DA}
\end{equation}
where $d_j(x_1,...,x_N)$ are the DAs. This relation is valid at the level of the bare operators.
If one assumes that the $\Omega_{N,\beta}^{(i)}$ functions are normalized to unity, i.e.,
\begin{equation}
\begin{split}
&\int \left[ d^2 \boldsymbol{k}_{\perp} \right]_N \psi^{(i)}\left(1, 2, \dots , N\right) = \phi^{(i)}_N(x_1,...,x_N) \int \left[ d^2 \boldsymbol{k}_{\perp} \right]_N \Omega^{(i)}_{N,\beta}\left(x_1, \boldsymbol{k}_{\perp 1}, x_2, \boldsymbol{k}_{\perp 2}, \dots,  x_N, \boldsymbol{k}_{\perp N} \right) = \phi^{(i)}_N(x_1,...,x_N)
\end{split}
\end{equation}
then we have the identification $\phi^{(i)}_N= \sum_{j} a_{ij} d_j$.
Notice that we can always impose that normalization to the $\Omega^{(i)}_{N,\beta}$ functions
if we do not assume specific boundary conditions for the $\phi$ functions.
In fact, given  Eq.~\eqref{Fundamental_LFWA_to_DA}, we can always introduce $\phi^{(i)}_N= \sum_{j} a_{ij} d_j$ and decompose 
\begin{equation}
\psi^{(i)}_N = \phi^{(i)}_N(x_1,...,x_N) \ta \frac{\psi^{(i)}_N(x_1,\boldsymbol{k}_{\perp 1},...,x_N,\boldsymbol{k}_{\perp N})}{\phi^{(i)}_N(x_1,...,x_N)}\tc,
\end{equation}
where the function in brackets can be named $\Omega_{N,\beta}^{(i)}$ and obviously satisfies the normalization condition 
\begin{equation}
\int \left[ d^2 \boldsymbol{k}_{\perp} \right]_N \Omega^{(i)}_{N,\beta} = 1.
\label{norm-omega}
\end{equation}
In the explicit construction of the LFWAs, we make a few assumptions.  First, we take the same  $\Omega_{N,\beta}^{(i)}$ function for each $N$ parton state,  independently on the different helicity structure of the  $N$ partons.
Specifically, we impose 
\begin{equation}
\Omega_{4,q\bar{q}gg}^{(1)} = \Omega_{4,q\bar{q}gg}^{(2)}, \quad \quad
\Omega_{4,q\bar{q}\mathcal{s}\bar{\mathcal{s}}}^{(1)} = \Omega_{4,q\bar{q}\mathcal{s}\bar{\mathcal{s}}}^{(2)} = \Omega_{4,q\bar{q}\mathcal{s}\bar{\mathcal{s}}}^{(3)}.
\end{equation}
This assumption is motivated by the fact that the  different $ \Omega^{(i)}_{N, \beta}$  functions for a given $N,\beta$ Fock state  contribute to the PDF only through the normalization  factor in Eq.~\eqref{eq:norm2} that is independent of the different partons' configuration. We also  verified that the  fit to the available experimental data for the form factor is not able to distinguish between different functional forms  for the  $ \Omega^{(i)}_{N, \beta}$ of a given $N,\beta$ Fock state. Accordingly, hereafter we can omit the the label $(i)$ in the $\Omega_{N,\beta}^{(i)}$ functions.
Moreover, we adopt an analogous simplification for the longitudinal-momentum dependence, as discussed in Sec.~\ref{Sec_PDFs}.

\subsubsection{Model for the transverse-momentum dependence}
For the functions $\Omega_{N,\beta}$ in Eq.~\eqref{Eq_Factorized_LFWA}, 
we adopt the Brodsky-Lepage-Huang~\cite{BHL} prescription which gives
\begin{equation}
    \label{Eq_Omega_functions}
    \Omega_{N,\beta} \left(x_1, \boldsymbol{k}_{\perp 1}, x_2, \boldsymbol{k}_{\perp 2}, \dots,  x_N, \boldsymbol{k}_{\perp N} \right) = \frac{(16 \pi^2 a_{\beta}^2)^{N-1}}{\displaystyle \prod_{i=1}^N x_i} \exp\left({-a^2_{\beta} \displaystyle \sum_{i=1}^N \frac{\boldsymbol{k}_{\perp i}^2}{x_i}}\right).
\end{equation}
The function in Eq.~\eqref{Eq_Omega_functions} satisfies the normalization condition in Eq.~\eqref{norm-omega}
and the following integral relation
\begin{equation}
    \label{Eq_Integrale_Omega_Quadro}
    \int \left[d^2 \boldsymbol{k}_{\perp} \right]_{N} \Omega_{N,\beta}^2 =  \frac{(8 \pi^2 a_{\beta}^2 )^{N-1}}{ \displaystyle \prod_{i=1}^N x_i}.
\end{equation}

We therefore have  four free parameters $a_{\beta}$ for the transverse-momentum dependent part of the LFWAs,  given by
\begin{eqnarray*}
&a_{q\bar q} \  \text{for the $q\bar{q}$ state},  & a_{q\bar q g} \ \text{for the $q\bar{q}g$ state}, \\
& a_{q\bar q gg} \ \text{for the state $q\bar{q}gg$ and }\quad \quad  & a_{q\bar q\mathcal{s}\bar{\mathcal{s}}
} \ \text{for the $q\bar{q}\mathcal{s}\bar{\mathcal{s}}$ state.}
\end{eqnarray*}
Let us denote the set of these parameters as $\bm{A} = \{a_{q\bar q},a_{q\bar q g},a_{q\bar q gg},a_{q\bar q \mathcal{s}\bar{\mathcal{s}}}\}$.

The model~\eqref{Eq_Omega_functions}  suffers from a minor inconvenience: either the DAs  or the PDFs   have some dependence on the transverse parameters $a_\beta$. This means that there is trace of the transverse structure in the collinear part. In principle, this is not an issue, since the dependence can be shown to be just a normalization factor. It is however an unwelcome feature in practical applications. For our purposes, it is important to avoid any dependence in the PDFs from the $a_{\beta}$ parameters. We therefore modify the model for $\Omega_{N,\beta}$ to read
\begin{equation}
\label{Eq_Omega_functions_v2}
\Omega_{N,\beta} \left(x_1, \boldsymbol{k}_{\perp 1}, x_2, \boldsymbol{k}_{\perp 2}, \dots,  x_N, \boldsymbol{k}_{\perp N} \right) = \frac{\left(4\sqrt{2} \pi a_{\beta}\right)^{N-1}}{\displaystyle \prod_{i=1}^N x_i} \exp\left({-a^2_{\beta} \displaystyle \sum_{i=1}^N \frac{\boldsymbol{k}_{\perp i}^2}{x_i}}\right).
\end{equation}

This implies for the normalizations
\begin{subequations}
\begin{eqnarray}
\int [d^2\boldsymbol{k}_\perp]_N \Omega_{N,\beta} &= \frac{1}{\left(2\sqrt{2}\pi a_{\beta}\right)^{N-1}}, \label{eq:norm1}\\
\int [d^2\boldsymbol{k}_\perp]_N \Omega^2_{N,\beta} &= \frac{1}{\prod_{i=1}^N x_i}.
\label{eq:norm2}
\end{eqnarray}
\end{subequations}
The net effect is to produce expressions for the PDFs without any dependence of $a_{\beta}$, while the DAs contain these parameters only as global normalization factors.
{We also note that the PDFs do not depend on the functional forms adopted for the $\Omega_{N,\beta}$ functions, once the normalization conditions in Eqs.~\eqref{eq:norm1} and \eqref{eq:norm2}  are imposed. Furthermore, the choice  in Eq.~\eqref{Eq_Omega_functions_v2} does not introduce a strong model dependence in the fit to the available data for the e.m. form factor: we verified that the fit results  are basically identical  if we adopt other functional forms, such as, for example, multipole-like functional forms or polynomials multiplied by a Gaussian.

\subsubsection{Model for the longitudinal momentum-fraction dependence}
\label{sec-long-parameters}
For the collinear part, our fundamental building block is the asymptotic expansion of the $x$-dependence of the leading-twist DAs (lowest conformal spin representation of the collinear conformal subgroup~\cite{Braun:1989iv}), which for the generic $N$-parton state reads
\[
\prod_{i=1}^N x_i^{2j_i-1},
\]
where $j_i$ the conformal spin of the $i$-th parton, which is $j = 1$ for quarks and anti-quarks and $j=3/2$ for gluons.
\\
 We found that truncating the expansion to the asymptotic expressions for the DAs of the $q\bar q$ and $q\bar q g$ components, it leads to poor results for the PDFs. Therefore, for these components, we included the first beyond-asymptotic term in the expansion. For the $q\bar q$ state, we modified also the asymptotic expansion, by assuming a variable exponent for the longitudinal momentum fractions to be fitted to data. 
We also included the first term orthogonal to this modified asymptotic DA, which is uniquely determined as the lowest-degree non trivial polynomial in the momentum fractions that is orthogonal to the asymptotic state when integrated with the two-dimensional simplex measure. This is exactly the expansion in  Gegenbauer polynomials with variable dimensionality, which has typically a faster convergence than the expansion with fixed dimensionality~\cite{Chang:2013pq}. 
For the remaining Fock states, we found that the variable exponents do not change significantly the results of the fit, and therefore we did not introduce these additional parameters.  Moreover, for the sea-quark contribution, we are not able to distinguish  the three LFWAs 
 $\phi^{(1)}_{q\bar{q}\mathcal{s}\bar{\mathcal{s}}}$, $\phi^{(2)}_{q\bar{q}\mathcal{s}\bar{\mathcal{s}}}$, and $\phi^{(3)}_{q\bar{q}\mathcal{s}\bar{\mathcal{s}}}$, as long as we consider unpolarized PDFs with the extant database. We then assume that these three LFWAs share the same dependence on the
longitudinal momentum fractions, though with a different normalization factor.
The situation is different for the two-gluon Fock component, for which the LFWA $\phi^{(2)}_{q\bar{q}gg}(x_1,x_2,x_3,x_4)$ is antisymmetric in the last two arguments, whereas the function $\phi^{(1)}_{q\bar{q}gg}(x_1,x_2,x_3,x_4)$ is symmetric. The relative normalizations of the different Fock states are fixed by the requirement on the normalization of the pion state:
\begin{align}
    \label{Eq_Normalization_Pion_State}
    \langle{\pi(P')}|{\pi(P)}\rangle = 2(2\pi)^3 P^+ \delta\left(P'^+ - P^+\right)
 \delta\left(\boldsymbol{P}'_\perp - \boldsymbol{P}_\perp\right).\end{align}

Explicitly the model reads:
\begin{align}
\phi_{q\bar{q}}^{(1)}(x_1,x_2) &= N_{q\bar{q}}^{(1)} (x_1 x_2)^{\gamma_{q}} \left(1 + d_{q1} C_{2}^{(\gamma_q+1/2)}(x_1-x_2)\right)\notag\\
& =  N_{q\bar{q}}^{(1)} (x_1 x_2)^{\gamma_{q}} \left\{1 + d_{q1} (1+2\gamma_q)[1+\gamma_q(x_1-x_2)^2-6x_1x_2]\right\}, \label{phi-qq}\\
\phi_{q\bar{q}g}^{(1)}(x_1, x_2, x_3 ) &= N_{q\bar{q}g}^{(1)} x_1x_2x_{3}^{2}\left[1 + d_{g1} (3-7x_3)\right], \\
\phi^{(1)}_{q\bar{q}gg} &= N^{(1)}_{q\bar{q}gg} x_1x_2(x_3x_4)^2 ,\\
\phi^{(2)}_{q\bar{q}gg} &= N^{(2)}_{q\bar{q}gg} x_1x_2x_3x_4(x_3^2-x_4^2), \\
\phi^{(1)}_{q\bar{q}\mathcal{s}\bar{\mathcal{s}}} &= N^{(1)}_{q\bar{q}\mathcal{s}\bar{\mathcal{s}}} x_1x_2x_3x_4, \\
\phi^{(2)}_{q\bar{q}\mathcal{s}\bar{\mathcal{s}}} &= N^{(2)}_{q\bar{q}\mathcal{s}\bar{\mathcal{s}}} x_1x_2x_3x_4,\\
\phi^{(3)}_{q\bar{q}\mathcal{s}\bar{\mathcal{s}}} &= N^{(3)}_{q\bar{q}\mathcal{s}\bar{\mathcal{s}}} x_1x_2x_3x_4,\label{phi-qqss}
\end{align}
where $C^{\alpha}_j(x)$ are Gegenbauer polynomials.
In Eqs.~\eqref{phi-qq}-\eqref{phi-qqss},
the norms are given by
\begin{align}
N_{q\bar{q}}^{(1)}&= \cos (\alpha_1)4^{\gamma_q}\sqrt{\frac{\Gamma\ta\frac{5}{2}+2\gamma_q\tc}{\Gamma\ta\frac{1}{2}\tc \Gamma\ta 2\gamma_q\tc}}\left\{(4\gamma_q+3)[  1+4\gamma_q  +2 d_{q1} (1-\gamma_q)(1+2\gamma_{q})]  + d_{q1}^2(1 + 2 \gamma_{q})^2(3 + 4\gamma_{q} +3\gamma^2_q)\right\}^{-1/2} ,  \\
N_{q\bar{q}g}^{(1)}&= 6\sqrt{210}\sin(\alpha_1)\cos(\alpha_2)\left(18-18d_{g1} + 29 d^2_{g1} \right)^{-1/2} ,\\
N_{q\bar{q}gg}^{(1)}&= 2\sqrt{\frac{7}{6}}N_{q\bar{q}gg}^{(2)} =30\sqrt{77} \sin(\alpha_1)\sin(\alpha_2)\cos(\alpha_3), \\
N_{q\bar{q}\mathcal{s}\bar{\mathcal{s}}}^{(1)}&= 
N_{q\bar{q}\mathcal{s}\bar{\mathcal{s}}}^{(2)} =
\frac{1}{\sqrt{2}}N_{q\bar{q}\mathcal{s}\bar{\mathcal{s}}}^{(3)} =
2\sqrt{35}\sin(\alpha_1)\sin(\alpha_2)\sin(\alpha_3),\,\, \mbox{for} \, \mathcal{s}\bar{\mathcal{s}}=u\bar u,\, d\bar d,\, s\bar s,
\end{align}
where $\Gamma$ is the Euler Gamma function.

In summary, we have the following parameters for the longitudinal momentum fraction dependence:
\begin{align*}
&\alpha_{1},\alpha_2, \alpha_3, \  \text{for the relative normalization of the different Fock components}; \\
& \gamma_q, \text{ that is the exponent of the modified asymptotic $q\bar{q}$ LFWA}; \\
& d_{q1}, \text{ that is the relative normalization between the zeroth and first term of the expansion of the $q\bar{q}$ LFWA}; \\
& d_{g1}, \text{ that is the relative normalization between the zeroth and first term of the expansion of the $q\bar{q}g$ LFWA}. 
\end{align*}
Let us denote the set of these parameters as $\mathcal{X} = \{\alpha_{1},\alpha_2,\alpha_3,d_{q1},d_{g1},\gamma_q\}$.

\section{Collinear parton distribution functions}\label{Sec_PDFs}
In this section, we apply the model for the pion LFWFs outlined above to the extraction of the pion PDFs from existing measurements. 
\\
The model scale is fixed at $\mu_{0}^{2} = 0.7225$ GeV$^2$, well below the charm mass threshold $m_{c}^{2} = 2.04$ GeV$^2$.
Moreover, by neglecting electroweak corrections and quark masses, charge symmetry imposes $f_{1,\pi^+}^u=f_{1,\pi^+}^{\bar d}=f_{1,\pi^-}^{d}
=f_{1,\pi^-}^{\bar u}=2f_{1,\pi^0}^{u}=2f_{1,\pi^0}^{\bar u}=2f_{1,\pi^0}^{d}=2f_{1,\pi^0}^{\bar d}$.
In the following,
we will refer to distributions in positively charged pions, using the notation $f_1$.
Assuming  also a SU(3)-symmetric sea, i.e., $f_{1}^{u} = f_{1}^{\bar{d}} = f_{1}^{s} = f_{1}^{\bar{s}}$, we end up with three independent PDFs: the total valence contribution $f_{1}^{v}$, the total sea contribution $f_{1}^{S}$, given by
\begin{align}
    \nonumber f_{1}^{v}& = f_{1}^{u_{v}}-f_{1}^{d_{v}}  = \big(f_{1}^{u} - f_{1}^{\bar{u}}\big) - \big(f_{1}^{d} - f_{1}^{\bar{d}}\big) 
     = 2 f_{1}^{u_{v}},\\
   f_{1}^{S}& = 2f_{1}^{u} + 2 f_{1}^{\bar{d}} + f_{1}^{s} + f_{1}^{\bar{s}}= 6 f_{1}^{u},     \label{Eq_SU3Symmetry2}
\end{align}
and the gluon contribution $f_{1}^{g}$.
The model-independent expressions for the pion PDFs in terms of overlap of LFWAs are collected in App.~\ref{App_LFWA_PDFs}. With 
those expressions and the model for the LFWAs built in Sec.~\ref{Sec_Model},   we obtain the following parametrizations 
\begin{align}
\label{Eq_Our_Valence_pdf_bar}   f_{1}^{v}(x,\mu^2_0) &=     \mathcal{C}^{v}_{q\bar{q}} (x\bar{x})^{2 \gamma_q-1} \left[1 + d_{q1} (1 + 2 \gamma_q) \ta 1 + \gamma_q (x-\bar{x})^2 - 6 x \bar{x} \tc \right]^2\\
\nonumber    &+  \mathcal{C}^{v}_{q\bar{q}g}  x \bar{x}^5 \left[3 +18x d_{g1} -10\bar{x} d_{g1}   + 13 d^2_{g1} + 14 x d^2_{g1}  (x-4\bar{x})\right]\\
\nonumber   & +  \mathcal{C}^{v}_{q\bar{q}gg} x \bar{x}^9 + \sum_{\{\mathcal{s} \bar{\mathcal{s}}\}}\mathcal{C}_{q\bar{q}\{\mathcal{s}\bar{\mathcal{s}}\}}^{v}  x \bar{x}^5 , \\
\label{Eq_Our_Gluon_pdf_bar} f_{1}^{g}(x, \mu^2_0)& =     \mathcal{C}^{g}_{q\bar{q}g}   (x\bar{x})^3 \left[1 + d_{g1} (3 - 7 x)\right]^2 + \mathcal{C}^{g}_{q\bar{q}gg_1}  x^3 \bar{x}^7+\mathcal{C}^{g}_{q\bar{q}gg_2} x\bar{x}^5 (5 -20x-6x^2 + 52x^3 + 95 x^4), \\
f_{1}^{S}(x, \mu^2_0) &=  2 \sum_{\{\mathcal{s} \bar{\mathcal{s}}\}}\mathcal{C}_{q\bar{q}\{\mathcal{s}\bar{\mathcal{s}}\}}^{S}  x \bar{x}^5,
\label{Eq_Our_Sea_pdf_bar}
\end{align}
where $\bar{x} = 1-x$ and the $\mathcal C$ coefficients are given by 
\begin{align}
\mathcal{C}^{v}_{q\bar{q}} &= \ta 2N_{q\bar{q}}^{(1)}\tc^2,\quad \quad
\mathcal{C}^{v}_{q\bar{q}g} = \frac{1}{15} \ta N_{q\bar{q}g}^{(1)}\tc^2 ,  \qquad\,\, \mathcal{C}^{v}_{q\bar{q}gg} = \frac{1}{315} \ta N_{q\bar{q}gg}^{(1)}\tc^2,\quad  \mathcal{C}_{q\bar{q}\{\mathcal{s} \bar{\mathcal{s}\}}}^{v}  =  \frac{1}{5}\ta N_{q\bar{q}\mathcal{s}\bar{\mathcal{s}}}^{(1)}\tc^2,\\
\mathcal{C}^{g}_{q\bar{q}g} &= \frac{1}{3}\ta N_{q\bar{q}g}^{(1)}\tc^2,\quad
\mathcal{C}^{g}_{q\bar{q}gg_1} =\frac{2}{105}\ta N_{q\bar{q}gg}^{(1)}\tc^2,\quad
\mathcal{C}^{g}_{q\bar{q}gg_2} = \frac{1}{4410}\ta N_{q\bar{q}gg}^{(1)}\tc^2,\\
\mathcal{C}^{S}_{q\bar{q}\{\mathcal{s} \bar{\mathcal{s}}\}} &= \frac{1}{10}\ta N_{q\bar{q}\mathcal{s}\bar{\mathcal{s}}}^{(1)}\tc^2 .
\end{align}
The valence number and momentum sum rules are guaranteed by construction, i.e.,
 \begin{align}
    \label{Sum_rules}
    \int_{0}^{1} f_{1}^v(x) dx  = 2, \quad \quad
    \int_{0}^{1} x \left( f_{1}^v(x) + f_{1}^g(x) + f_{1}^S(x)  \right)dx  = 1.
\end{align}

\subsection{Fit procedure}
The PDF fit of the cross-section experimental data has been performed by using the open-source tool xFitter~\cite{Alekhin:2014irh}. Technical details about the fit set-up can be found in~\cite{Novikov:2020snp}. Here we report  only the essential information necessary to reproduce our fit results. 
 In our analysis, we consider DY data  from the NA10~\cite{NA10:1985ibr} and E615~\cite{Conway:1989fs} experiments and  prompt photon production data (WA70)~\cite{WA70:1987bai}. The DY data  have been obtained by studying the scattering of a $\pi^-$ beam off a tungsten target, with the pion beam energy $E_{\pi} = 194 \text{ and } 286 \text{ GeV}$ in the NA10 experiment, and $E_{\pi} = 252 \text{ GeV}$ in the E615 experiment. Instead, the prompt photon production data of the WA70 experiment have been measured by using $\pi^{\pm}$ beams, with $E_{\pi} = 280 \text{ GeV}$, on a proton target.
In order to avoid the $J/\psi$ and $\Upsilon$ resonances and the lower edges of phase space, we apply the following cuts:  $4.16 \text{ GeV }< \mu < 7.68 \text{ GeV}$, and $x_F \geq 0 $, where the Feynman variable $x_F$ is defined as $x_F=x_0^\pi-x_0^A$ with $x_0^{\pi(A)}$  the minimum momentum fraction of the active parton in  the pion (nucleus) to
produce the lepton pair in the final state.
The number of data points after the cuts are 91  for the E615 set and 70 for the NA10 set. Including also the prompt-photon data, our database consists of a total number of 260 points.
\\

The minimization function is defined as:
\begin{align}
    \label{Eq_Chi_squared_PDFs}
    \chi^2 = \sum_{i}\frac{\left(E_i - \tilde{t}_i\right)^2}{\left(\delta_{i}^{\text{syst}}\right)^2 + \left(\sqrt{\frac{\tilde{t}_i}{d_i}} \delta_{i}^{\text{stat}}\right)^2} + \sum_{\alpha}b_{\alpha}^2,
\end{align}
where the index $i$ runs over the data points and $\alpha$ is the index of the source of correlated error.
In Eq.~\eqref{Eq_Chi_squared_PDFs}
$E_i$ are the measured cross sections with the corresponding systematic and statistical uncertainties $\delta_{i}^{\text{syst}}$ and $\delta_{i}^{\text{stat}}$, respectively. Moreover,
 $\tilde{t}_i = t_i\left(1 - \sum_{\alpha} \zeta_{i \alpha} b_{\alpha}\right)$ are the theory predictions $t_i$  corrected for the correlated shifts, obtained by
 taking into account the relative coefficient $\zeta_{i\alpha}$ of the influence of the correlated error source $\alpha$ on the data point $i$ and the nuisance parameter $b_\alpha$. The nuisance parameters  are included in the minimization along with the PDF parameters
  and contribute to the $\chi^2$ via the penalty term $\sum_{\alpha} b_{\alpha}^2$~\cite{Novikov:2020snp}. 
  
\subsection{Fit results}
The model parameters to be fitted are the 6 collinear coefficients of the pion PDFs  described in Sec.~\ref{sec-long-parameters}.
In addition to the initial scale  $\mu_{0}$, we fixed the factorization scale $\mu_F$ and the renormalization scale $\mu_R$ to $\mu_F=\mu_R=0.8$ GeV. 
The minimization of the function~\eqref{Eq_Chi_squared_PDFs} is performed by using MINUIT~\cite{James:1975dr}.
For the averaged value of the fitted parameters we find
\[
\begin{split}
\langle{d_{q1}}\rangle     &= -0.142,  \quad \langle{\gamma_{q}}\rangle = 0.639, \quad  \langle{d_{g1}}\rangle     = 111.386, \\
\langle{\alpha_1}\rangle        &= 0.816,   \quad\,\,\,\;
\langle{\alpha_2}\rangle        = 1.364,   \quad \;
\langle{\alpha_3}\rangle       = 0.554. 
\end{split}
\]
The  reduced chi-squared from a single minimization is $\hat{\chi}^2/N_{d.o.f.} = 0.88$ for the number of degrees of freedom $N_{d.o.f.} = 260-6=254$.

We  also repeated the fit by varying the value of the initial scale around $\mu_0=0.85$ GeV and we did not observe significant variations in the quality of the fit (the chi-squared varied by less then $1\%$ when changing $\mu_0$ in the range $[0.65, 1.05]$ GeV).
The error analysis is performed with the bootstrap method, by fitting an ensemble of 1000 replicas of experimental data varied  by using a random gaussian shift both for the statistic and systematic uncertainties. 
Furthermore, we took into account  the effects of variations of the factorization and renormalization scales by changing the values of $\mu_F$ and $\mu_R$ replica by replica. In particular, the value of  $\mu_F$ has been randomly generated from a uniform distribution in the range $\left[\mu_0/2, \mu_0\right]$, while  $\mu_R$ was varied in the range $\left[\mu_{F}, 2\mu_0\right]$, i.e, we explored the region of $\mu_0/2 \leq \mu_F \leq \mu_0$ and $\mu_0/2 \leq \mu_R \leq 2 \mu_0$.

 \begin{figure}[t]
    \centering
    \includegraphics[width=0.49\textwidth]{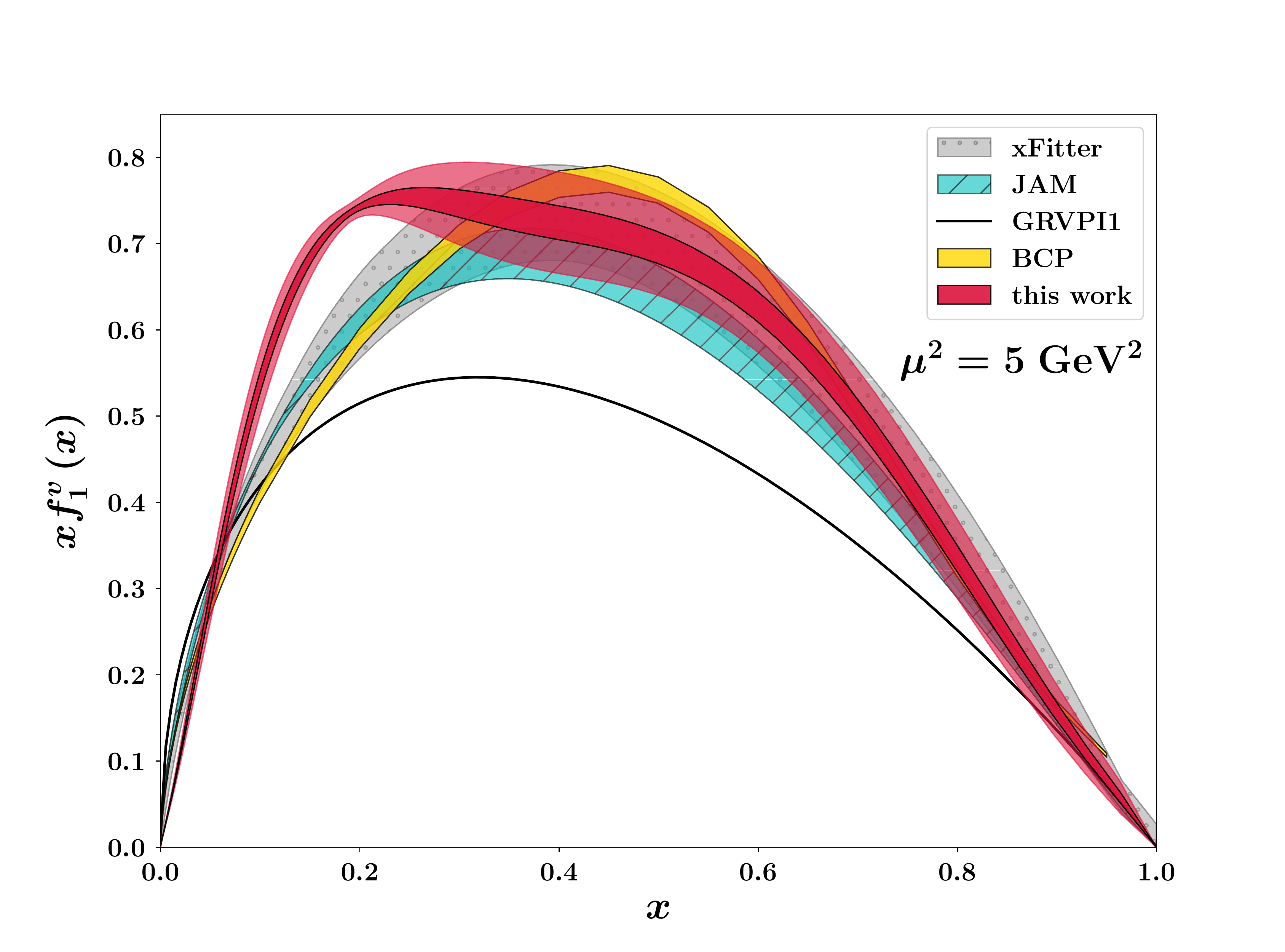}
    \\
    \includegraphics[width=0.49\textwidth]{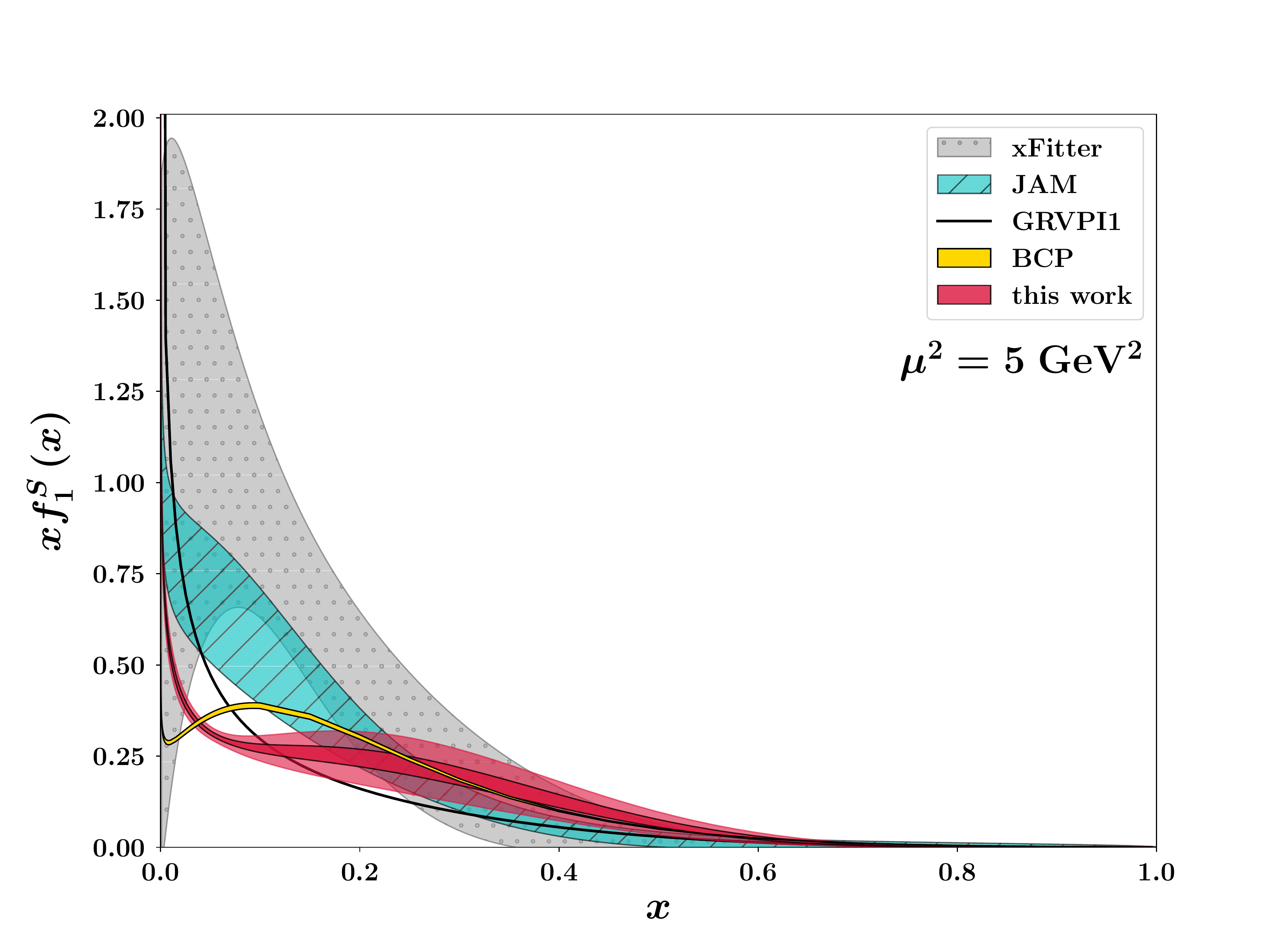}
    \includegraphics[width=0.49\textwidth]{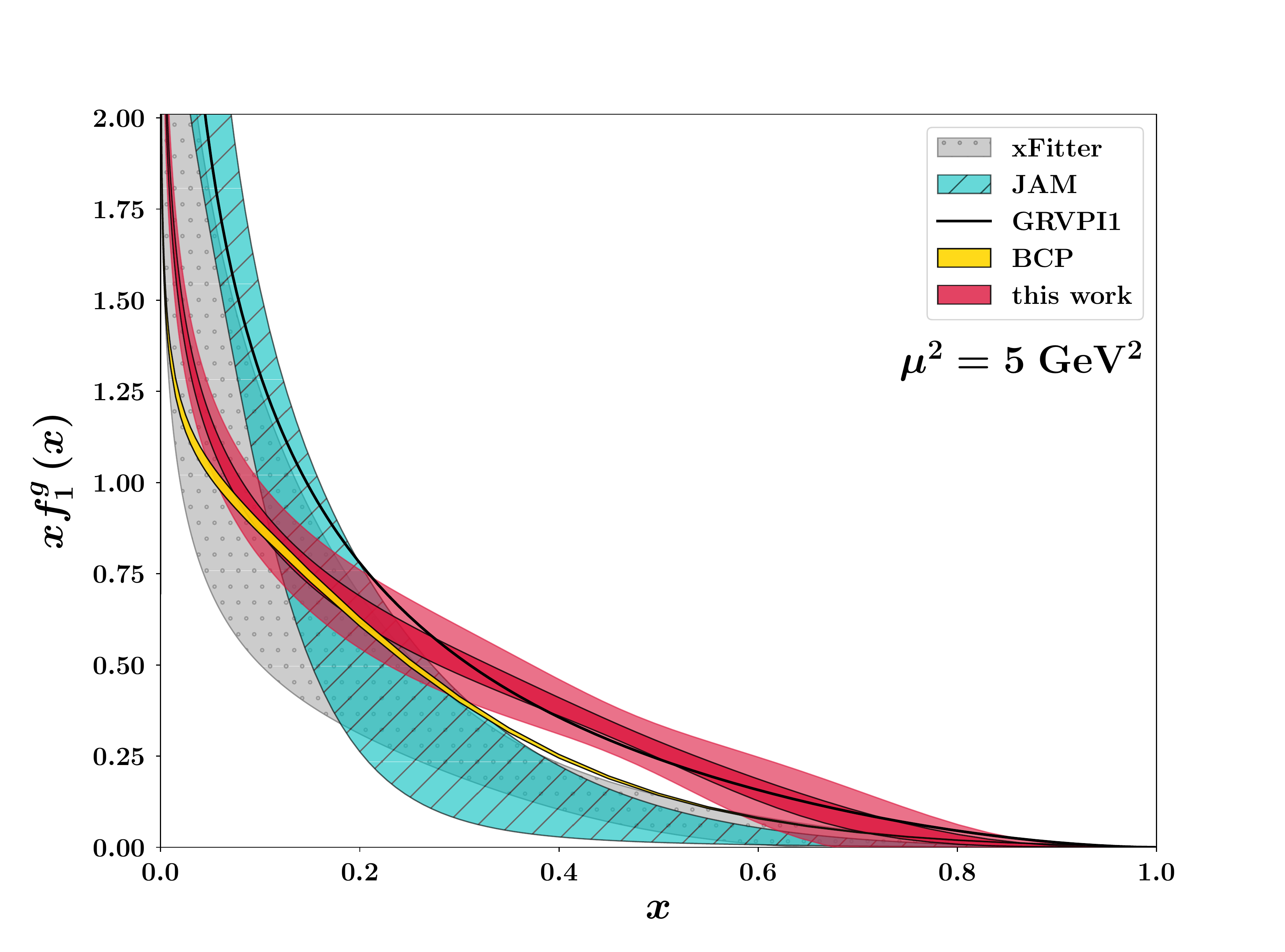}
    \caption{ $xf_{1}$ as function of $x$  for the  total valence (upper panel), total sea  (left panel in the bottom) and gluon (right panel in the bottom) contributions  at $\mu^2=5$ GeV$^2$. The light (dark) red bands show the results of this work with the 3$\sigma$  (1$\sigma$) uncertainty in comparison with the results from the JAM collaboration~\cite{Barry:2021osv} (light blue bands), the analysis of xFitter collaboration~\cite{Novikov:2020snp} (grey bands),  the BCP fit  of Ref.~\cite{Bourrely:2022mjf} (yellow bands) and  the GRVPI1 fit~\cite{Gluck:1991ey} (solid black  curves). 
    \label{Fig_All_PDFs_5}}
\end{figure}
 In Fig.~\ref{Fig_All_PDFs_5}  we show the results for the pion PDFs at the scales of $\mu^2=5$ GeV$^2$. The light and dark red uncertainty bands are our fit results, corresponding, respectively, to $3\sigma$ ($99.7$\%) and $1\sigma$ ($68$\%) confidence level (CL). They are compared with  the extractions of pion PDFs from other studies. The solid black  curves correspond  to best fit  of  the analysis in Ref.~\cite{Gluck:1991ey} (GRVPI1 solution) and the grey bands refer to the results of the xFitter collaboration~\cite{Novikov:2020snp}. These analyses and our work were
based on the same measurements, with small variations in the database due to different  kinematical cuts (we refer to the original works for details).
The analysis from the JAM collaboration~\cite{Barry:2021osv} is shown by the light blue bands and includes both the
DY data and the leading-neutron tagged electroproduction data, taking into account also threshold resummation on DY cross sections at next-to leading log accuracy.
The yellow bands show
a new analysis by Bourrely, Chang and Peng (BCP) in the framework of the statistical model~\cite{Bourrely:2022mjf} which extended the database considered in a previous work~\cite{Bourrely:2020izp} to include  $J/\psi$ production data.
Overall, the modern analyses give compatible results within the relative error bands. The agreement is better  for the valence and sea contributions at larger $x$ and for the gluon PDF in the small $x$ region. The difference in shape of our results for the valence  PDF in the region $0.05<x<0.2$, can be ascribed to a strong correlations between the valence PDF at small $x$ and the gluon PDF at large $x$.
This is peculiar to the LFWF approach. From the explicit expressions for the PDFs in Eqs.~\eqref{Eq_Our_Valence_pdf_bar}-\eqref{Eq_Our_Sea_pdf_bar}, we notice that the valence PDF receives contributions from all  Fock states. In particular, the low-$x$ behavior of the valence PDF is influenced by the high-$x$ behavior of the other PDFs. These spurious correlations tend to lessen 
when the expansion in the Fock space spans a large number of Fock components. However, one has to admit that the extension of the present formalism at higher-order Fock components may become quite cumbersome.
\begin{figure}[t]
    \centering
    \centering
    \includegraphics[width=0.49\textwidth]{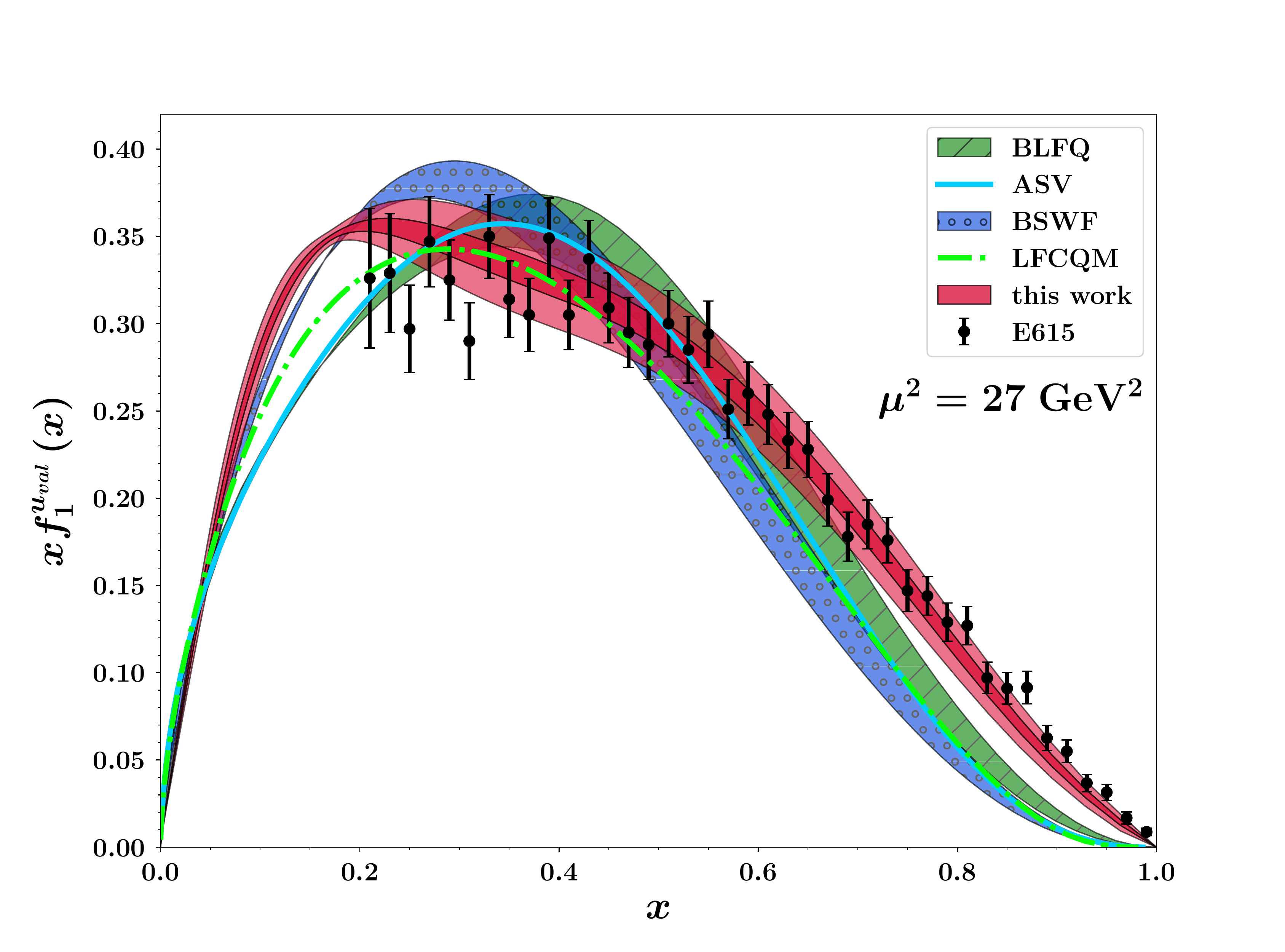}
    \\
     \includegraphics[width=0.49\textwidth]{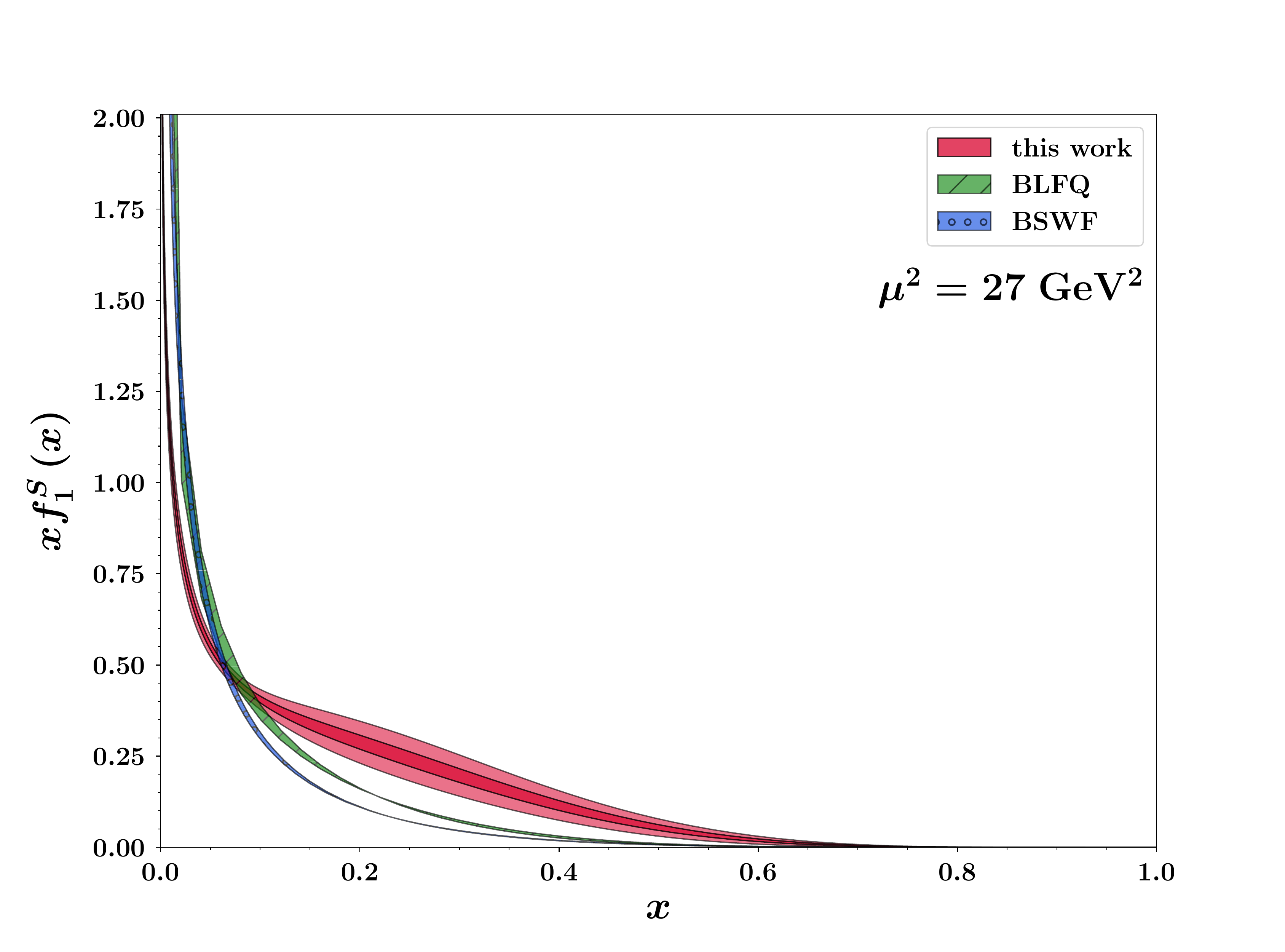}
    \includegraphics[width=0.49\textwidth]{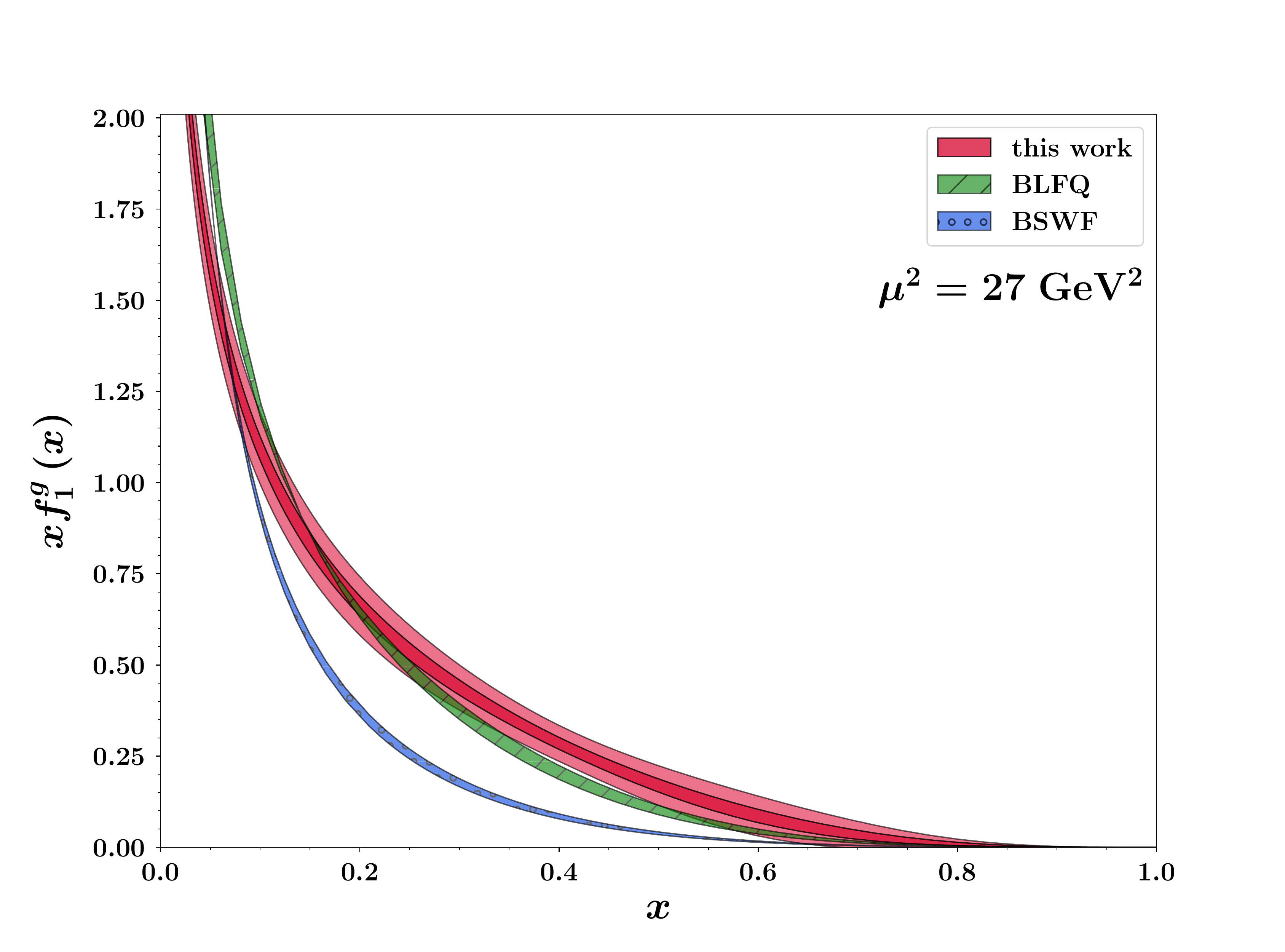}
    \caption{$xf_{1}$ as function of $x$  for the valence $u$ quark (upper panel), total sea  (left panel in the bottom) and gluon (right panel in the bottom) contributions  at $\mu^2=27$ GeV$^2$. The light (dark) red bands show the results of this work with the 3$\sigma$  (1$\sigma$) uncertainty in comparison with the 
   the BSWF results  from Ref.~\cite{Cui:2020tdf} (blue bands)  and the analysis from the BLFQ collaboration~\cite{Lan:2021wok} (green bands). For the valence $u$ quark contribution we show also the  ASV parametrization of Ref.~\cite{Aicher:2010cb} (cyan solid curve),  and the results within the LFCQM of Ref.~\cite{Pasquini:2014ppa} (green dashed-dotted curve), while the  data are from the E615 experiment~\cite{Conway:1989fs}.
   \label{Fig_All_PDFs_27}}
\end{figure}

In Fig.~\ref{Fig_All_PDFs_27}, we show our results at $\mu^2=27$ GeV$^2$  with light (dark) red bands corresponding to $3\sigma$ ($1\sigma$) CL. The  $u$-quark valence PDF  is in very good agreement with the  extraction of the E615 experiment~\cite{Conway:1989fs} which neglected threshold resummation effects as we did in our fit procedure.
However, in a seminal paper 
Aicher, Sch\"afer, and
Vogelsang (ASV)~\cite{Aicher:2010cb} found that corrections from threshold
resummation can  significantly modify  the large-$x$ asymptotic behavior of the valence quark contribution, as shown  by the solid cyan curve in Fig.~\ref{Fig_All_PDFs_27}\footnote{The ASV analysis focused mainly on the fit of the valence PDF, using  parametrizations for the gluon and sea contribution from other works~\cite{Gluck:1999xe} which are not shown here.}.
This large-$x$ behavior is reproduced very well from the light-front constituent quark model (LFQCM) predictions of Ref.~\cite{Pasquini:2014ppa} (green dashed-dotted curve),  which considered only the $q\bar q$ Fock component at the hadronic scale and applied NLO evolution to the relevant experimental scale.
Analogously, the results of the basis light-front quantization (BLFQ) collaboration~\cite{Lan:2021wok} within a light-front model including the $q\bar q$ and $q\bar q g$ Fock components and the 
study of Ref.~\cite{Ding:2019lwe}  with Bethe-Salpeter wavefunctions (BSWF) for the $q\bar q$ state are consistent with the behavior at large $x$ inferred from the analyses with the threshold 
resummation.
In Fig.~\ref{Fig_All_PDFs_27}, we also compare the outcome of our study for the sea and gluon contributions with the BLFQ and BSWF results. While our PDF parametrizations take into account non-perturbative sea and gluon contributions at the initial hadronic scale, the BSWF approach generates both the sea and gluon contributions  solely from the scale evolution, and the BLFQ model includes only a dynamical gluon contribution at the initial scale and generates the sea PDF perturbatively. 
One can clearly appreciate the effects of  non-perturbative sea and gluon contributions  that result at $x\ge 0.1$ in larger sea PDF in our analysis, and larger gluon PDF in our and BLFQ analyses. We also observe a steeper rise at lower values of $x$ in the BLFQ and BSWF models than in our results for the gluon and sea PDFs. Despite the different shapes of the  PDFs in our fit and the BSWF and BLFQ analyses, the first moments of the PDFs, defined as $\int {\rm d}x \, x\, f_1$, are well compatible within the error bars, as shown in Table~\ref{Tab_Momenta}.
\renewcommand{\arraystretch}{1}
\begin{table}[H]
\begin{center}
\begin{tabular}{ccccc}
\hline \hline
 & $\mu^2$ GeV$^2$ & $\langle xf_{1}^{v}\rangle $ & $\langle xf_{1}^{S}\rangle $  & $\langle xf_{1}^{g}\rangle $    \\
\hline
JAM-Res~\cite{Barry:2021osv}
& $1.61$  
& $0.53\pm 0.02$
& $0.14\pm 0.04$
& $0.34\pm 0.06$
 \\
 \\
BLFQ~\cite{Lan:2021wok}
& $1.69$  
& $0.536$
& $0.069$
& $0.395$ 
 \\
 JAM~\cite{Barry:2018ort}
 &$1.69$
 &$0.54\pm 0.01$
 &$0.16\pm0.02$
 &$0.30\pm 0.02$\\
 xFitter~\cite{Novikov:2020snp}
 &$1.69$
 &$0.55\pm 0.06$
 &$0.26\pm0.15$
 &$0.19\pm0.16$\\
\rowcolor{tableGray} 
This work
 &$1.69$
 & $0.58 \pm 0.03$
 & $0.09\pm0.04$
 & $0.33\pm0.06$
 \\
 \\
 Latt1~\cite{Meyer:2007tm}
 & $4$
 &
 &
 &$0.37\pm 0.08\pm 0.12$
 \\
 Latt2~\cite{Shanahan:2018pib}
 &$4$
 &
 &
 &$0.61\pm 0.09$
 \\
 Latt3~\cite{Oehm:2018jvm}
&$ 4$
&$0.415\pm0.0212$
&
&
\\
Latt4~\cite{Joo:2019bzr}
&$4$
&$0.376\pm 0.112$
&
&
\\
Latt5~\cite{ExtendedTwistedMass:2021rdx}
&$4$
&  
& 
&$0.52\pm 0.11^{+0.02}_{-0.00}$
\\
BLFQ~\cite{Lan:2021wok}
&$ 4$
&$ 0.484$
&$ 0.094$
&$ 0.421$
\\
 BSWF~\cite{Cui:2020tdf}
 &$4$
 &$0.47\pm0.02$
 &$0.11\pm0.02$
 &$0.41\pm0.02$
 \\
 xFitter~\cite{Novikov:2020snp} 
 &$4$
 &$0.50\pm0.05$
 &$0.25\pm0.13$
 &$0.25\pm0.13$
 \\
\rowcolor{tableGray} This work
 &$4$
 &$0.52\pm0.03$
 &$0.11\pm0.03$
 &$0.37\pm0.05$
 \\
\\
 JAM~\cite{Barry:2018ort}
 &$5$
 &$0.48\pm0.01$
 &$0.17\pm0.01$
 &$0.35\pm0.02$
 \\
 xFitter~\cite{Novikov:2020snp}
 &$5$
 &$0.49\pm 0.05$
 &$0.25\pm 0.12$
 &$0.26\pm 0.13$
 \\
  BCP~\cite{Bourrely:2022mjf}
  &$5$
  &$0.50\pm 0.01$
  &$0.19\pm 0.012$
  &$0.31\pm 0.002$
  \\
\rowcolor{tableGray} This work
 &$5$
 &$0.51\pm0.03$
  &$0.12\pm0.03$
 &$0.37\pm0.05$
  \\ \\
  BLFQ~\cite{Lan:2021wok}
  &$10$
  & $ 0.446$
  &$ 0.115$
  &$ 0.439$
  \\
 JAM~\cite{Barry:2018ort}  
&$10$
 &$0.44\pm 0.01 $
 &$0.19\pm0.01$
 &$0.37\pm 0.02$\\ 
 xFitter~\cite{Novikov:2020snp}
 &$10$
 &$0.46\pm 0.02$
 &$0.22\pm 0.08$
 &$0.31\pm 0.06$
 \\
 BCP~\cite{Bourrely:2022mjf}
 &$10$
 &$0.48\pm0.08$
 &$0.21\pm0.012$
 &$0.33\pm0.015$
 \\
\rowcolor{tableGray} This work
 &$10$
 &$0.48\pm0.03$
 &$0.13\pm0.02$
 &$0.39\pm0.05$
 \\
 \\
 Latt4~\cite{Joo:2019bzr}
&$27$
&$0.330\pm 0.018$
&
&
\\
 BLFQ~\cite{Lan:2021wok}
 &$27$
 & $0.414$
 &$ 0.132$
 &$ 0.451$
 \\
 BSWF~\cite{Cui:2020tdf}
 &$27$
 &$0.41\pm0.04$
 &$0.14\pm0.02$
 &$0.45\pm0.02$
 \\
xFitter~\cite{Novikov:2020snp}
 &$27$
 &$0.42\pm0.04$
 &$0.25\pm0.10$
 &$0.32\pm0.10$\\
 \rowcolor{tableGray}This work
 &$27$
 &$0.45\pm0.02$
 &$0.15\pm0.02$
 &$0.40\pm0.04$
 \\
 \hline
\hline
\end{tabular}
\end{center}
\caption{
Results from this work (grey rows) for the momentum fractions of the pion carried by the valence, sea and gluon PDFs at different scales $\mu^2$, in comparison with other phenomenological extractions, model calculations and lattice-QCD analyses.   \label{Tab_Momenta}}
\end{table}
In the same table, we also collect the results of  other studies at different scales.
We observe   that at the initial scale $\mu^2=1.69$ GeV$^2$ of  the xFitter and JAM analyses, the values for the gluon are larger in our model and JAM input than in xFitter,  although they are still consistent within the error bar.  The same trend remains at higher scales after evolution.
 Comparing our results with the recent BCP extractions, we find a remarkable agreement for the valence moments, while our values for the sea contribution are  smaller, mainly because of the different behavior of the sea PDFs at 
 $x\lesssim 0.1$ as shown, for example, in Fig.~\ref{Fig_All_PDFs_5} for the results at $\mu^2$=5 GeV$^2$.
The lattice calculations  for the valence contribution obtained systematically smaller values than the phenomenological analyses, while the lattice values for the gluon
changed
significantly going  from the analysis of Ref.~\cite{Meyer:2007tm} using quenched QCD and a large 800 MeV pion mass to the  study of Ref.~\cite{Shanahan:2018pib} with clover fermion action and 450 MeV pion mass. 
We also report the result for the gluon contribution of a recent calculation~\cite{ExtendedTwistedMass:2021rdx} which presents for the first time the decomposition into gluon and quark contributions.
In particular, they 
 calculated  the total $u+d$, $s$ and $c$ contributions, from which we can not reconstruct the separate valence and sea contributions.
Their results  for the sum over all quark flavors is $\sum_q \braket{x f_1^q} = 0.68 \pm 0.05^{+0.00}_{-0.03}$, while the 
the sum of all contributions
amounts to $1.20 \pm 0.13^{+0.00}_{-0.03}$, compatible with the
expected value of 1 within two sigma.
There exists also a new lattice study of the $x$-dependence of the gluon PDF at   $\mu^2$=4 GeV$^2$~\cite{Fan:2021bcr}. It reports the results for $xf_{1}^g(x)$  normalized to unity and gives indications that future lattice studies 
with improved precision and systematic control may help to provide best determinations of the gluon content in the pion when combined in global-fit analyses.

\section{Electromagnetic Form Factor}
\label{Section_FF}
As the PDFs, the e.m. form factor (FF) can be written in terms of an overlap of LFWAs, see App.~\ref{App_LFWA_FFs}. The non-diagonal matrix elements prevent one from using Eq.~\eqref{Eq_Integrale_Omega_Quadro} in the computation.
In principle one can obtain full analytical expression for the form factor, although it is long and rather uninformative. We present here the model result in terms of the contributions of the different Fock states and in the implicit integral form, which allows for compact expressions, i.e.,
\begin{align}
\label{Fpi_sum_FockComp}
F_{\pi}(Q^2)&=F_{\pi,q\bar{q}}(Q^2)+F_{\pi,q\bar{q}g}(Q^2)+
F_{\pi,q\bar{q}gg}(Q^2)
+\sum_{\{\mathcal{s} \bar{\mathcal{s}}\}}F_{\pi,q\bar{q}\{\mathcal{s}\bar{\mathcal{s}}\}}(Q^2),
\end{align}
with
\begin{align}
F_{\pi,q\bar{q}}(Q^2) &= \frac{1}{2}\mathcal{C}^v_{q\bar{q}}\int_0^1 dx \  (x\bar{x})^{2 \gamma_q-1} \left[1 + d_{q1} (1 + 2 \gamma_q) \ta 1 + \gamma_q (x-\bar{x})^2 - 6 x \bar{x} \tc \right]^2 \exp\left[ -a^2_{q\bar q}Q^2\frac{\bar{x}}{2x}\right],\nonumber
\\
F_{\pi,q\bar{q}g}(Q^2) &= \frac{1}{2}\mathcal{C}^v_{q\bar{q}g}\int_0^1 dx \ x\bar{x}^5 \ta 3+18xd_{g1}-10\bar{x}d_{g1}+13d_{g1}^2+14xd_{g1}^2(x-4\bar{x})\tc\exp\left[ -a^2_{q\bar q g}Q^2\frac{\bar{x}}{2x}\right],\nonumber
\\
F_{\pi,q\bar{q}gg}(Q^2) &= \frac{1}{2}\mathcal{C}^v_{q\bar{q}gg}  \int_0^1 dx \ x\bar{x}^9 \exp\left[ -a^2_{q\bar q gg}Q^2\frac{\bar{x}}{2x}\right],\nonumber
\\
F_{\pi,q\bar{q}\{\mathcal{s}\bar{\mathcal{s}}\}}(Q^2) &= \frac{1}{2} \mathcal{C}^v_{q\bar{q}\mathcal{s}\bar{\mathcal{s}}}\int_0^1 dx \ x\bar{x}^5 \exp\left[ -a^2_{q\bar q \mathcal{s}\bar{\mathcal{s}}}Q^2\frac{\bar{x}}{2x}\right],
\label{Fpi_sum_FockComp2}
\end{align}
where $Q^2=-q^2>0$ and $q=p'-p$  the four-momentum transfer.
The form factor is normalized as $F_\pi(Q^2=0)=1$, consistently with the valence sum rule.
The collinear parameters $\mathcal{X}$ of the LFWAs have been determined by the fit of the pion PDF. The only free parameters in the fit of the pion FF are those entering in the $\Omega_{N,\beta}$ functions of Eq.~\eqref{Eq_Omega_functions}, i.e the set $\bm{A}$.

\subsection{Fit procedure}
The available experimental data of the pion 
e.m. form factor come from different extractions exploring various kinematic regions. The CERN measurements of $\pi$-$e$ scattering experiments~\cite{NA7:1986vav} provide data for the square of the pion e.m. form factor  in the range $0.015$  GeV$^2 \leq Q^2 \leq 0.253$ GeV$^2$. The extension to larger values of $Q^2$ requires the use of pion electroproduction from a nucleon target (Sullivan process~\cite{Sullivan}). This process has been exploited for the extraction of $F_{\pi}$ at JLab ~\cite{JeffersonLabFpi:2000nlc,JeffersonLab:2008jve,JeffersonLabFpi-2:2006ysh}  and DESY~\cite{Bebek:1977pe}. 
Combining all the data sets, we have 100 experimental points in a $Q^2$ range from $0.015$ GeV$^2$ to $9.77$ GeV$^2$.

As in the case of the PDF, we use the bootstrap replica method to propagate the experimental uncertainties to our final results. A detailed explanation on the inner workings of the method and examples for its application to  extractions involving multiple data sets with systematic errors can be found in Ref.~\cite{Pedroni:2019dlg}.
Moreover, as detailed in Ref.~\cite{Pedroni:2019dlg}, the bootstrap technique is  useful also when one wants to propagate the uncertainties associated to parameters of the model that are not directly free fitting variables. In our case, these are the parameters entering
the collinear part of the LFWAs.
The general logic of the fit procedure is as follows: from the PDF extraction, we obtained a set of $n=1000$ vectors of parameters, one for each of the replica of the data used for PDF extraction. Then, for each bootstrap cycle in the form-factor fit, we generate a replica of the data and sample (uniformly) a vector of PDF parameters from the set generated in the PDF fit. We then perform the fit of the form factor  computed with the sampled PDF parameters and the free $a_\beta$ parameters. The minimization function is the bootstrap $\chi^2_k$ for the $k$-th bootstrap replica, i.e.,
\[
\chi^2_k = \sum_{j=1}^{n_{\text{set}}} \sum_{i=1}^{n_{d,\text{set}_j}} \frac{(F^{\mathcal{X}_k}_\pi(Q^2)- d_{ijk})^2}{\sigma_{ijk}^2},
\]
where  $n_{d,\text{set}_j}$ is the number of data in the $j$-th dataset,  $n_{\text{set}}$ is the total number of dataset and $F^{\mathcal{X}_k}_\pi(Q^2)$ 
is the pion e.m. form factor computed using the $k$-th uniformly sampled vector  of collinear parameters $\mathcal{X}_k$.
The bootstrap quantities are defined as
\[
d_{ijk} = (1+\delta_{ik})\ta d_{ij} + \sigma_{ijk} \tc = (1+\delta_{ik})\ta d_{ij} + r_{ijk}\sigma_{ij} \tc,
\]
where $d_{ij}$ and $\sigma_{ij}$ are the experimental data points and error, respectively, $\delta_{ik}$ is a random value extracted from the distribution for the systematic error of set $i$, and $r_{ijk}$ is a random number extracted from a normal distribution. 

After the minimization is performed for all bootstrap replicas, we can use the results to construct the multidimensional probability distribution for the parameters, and crucially extract the correlation coefficients for all pairs of parameters.

The set of transverse parameters $\bm{A}$ contains four distinct elements. However, due to the extremely small norm of the $q\bar{q}g$ state as obtained from the PDF fit, the form-factor fit is not sensitive to the value of $a_{q\bar{q}g}$. We will therefore exclude this parameter from the fit by fixing it arbitrarily to $1$ GeV$^{-1}$.
The specific value is inconsequential, since ultimately the corresponding contribution to the form factor is irrelevant.
\subsection{Fit results}

Including systematic uncertainties for different data sets and the sampling for the non-fitted parameters,  it implies that the probability distribution for the chi-squared can be substantially different from the  standard chi-squared distribution. To estimate the confidence level for the value of the single minimization $\hat{\chi}^2$, we perform a second bootstrap against fake data generated from the average value of parameters extracted from the `true' bootstrap run. The set of chi-squared obtained from this second bootstrap represents an estimation of the confidence interval (CI) for the chi-squared computed from a single minimization of our model, without replica of the data. Details on the theory behind this procedure are given in~\cite{Pedroni:2019dlg}.
We obtain for the reduced chi-squared $\hat{\chi}^2/N_{d.o.f.}$:
\[
\hat{\chi}^2/N_{d.o.f.} = 1.194, \quad \text{CI 68\%} = [0.890,1.204], \quad \text{CI 99\%} = [0.682, 1.593],
\]
for 
$N_{d.o.f.} = N-N_{\bm{A}} = 97$,
with the total number of data $N=100$ and the number of fitting parameters $N_{\bm{A}} = 3$. This shows that the single-minimization $\hat{\chi}^2$ is inside the 68\% confidence interval. 

For the minimized parameters, we find (in units of GeV$^{-2}$)
\[
a^2_{q\bar q} = 1.559, \quad a^2_{q\bar q gg} = 0.509, \quad a^2_{ q\bar q \mathcal{s}\bar{\mathcal{s}  }}= 0.796.
\]
\renewcommand{\arraystretch}{1.2}
\begin{table}[b]
    \centering
    \begin{tabular}{c|ccccccccc}
    & $a^2_{q\bar q}$ &  $a^2_{ q\bar q gg}$ & $a^2_{q\bar q \mathcal{s} \bar{  \mathcal{s}}}$ & $d_{1g}$ & $d_{1q}$ & $\gamma_q$ & $\alpha_1$ & $\alpha_2$ & $\alpha_3$ \\
    \hline
$a^2_{ q\bar q}$ &0.356 & -0.593 & -0.656 &0.067& -0.050& -0.056 &0.056 &-0.088 &-0.046 \\
$a^2_{q\bar q gg}$ &-0.593& 0.080 &0.484 &-0.381  & 0.069& 0.099 &-0.288 &0.599 &-0.161\\
$a^2_{q\bar q \mathcal{s} \bar{\mathcal{s}}}$ &-0.656& 0.484 &0.309 &-0.181  & 0.076& 0.081 &-0.177 &0.273 &-0.011\\
$d_{1g}$&0.067 &-0.381 &-0.181& 220.545&-0.260& -0.253& 0.277 &-0.495& 0.002\\
$d_{1q}$&-0.050& 0.069 &0.076 &-0.260  & 0.047& 0.982 &-0.108 &0.048 &0.245\\
$\gamma_q$&-0.056& 0.099 &0.081 &-0.253  & 0.982& 0.100 &-0.083 &0.079 &0.195\\
$\alpha_1$&0.056 &-0.288 &-0.177& 0.277  &-0.108& -0.083& 0.018 &-0.635& -0.605\\
$\alpha_2$&-0.088& 0.599 &0.273 &-0.495  & 0.048& 0.079 &-0.635 &0.202 &0.054\\
$\alpha_3$&-0.046& -0.161& -0.011& 0.002 & 0.245& 0.195 &-0.605 &0.054 &0.082\\
    \end{tabular}
    \caption{Full correlation table for the fitted parameters. On the diagonals are shown the standard deviations of the parameters. All the off-diagonal elements represent the linear correlation coefficients of the corresponding pair of parameters.
    }
    \label{Tab_res_FF_cov}
\end{table}
In Tab.~\ref{Tab_res_FF_cov} we show
the  correlation matrix of the parameters.
We notice a very large error for the  $d_{1g}$ parameter. This is because
the PDF fit prefers configurations with very small norm for the $q\bar{q}g$ component, and therefore it is almost insensitive to the $d_{1g}$ parameter. 
Moreover,
the strong correlations between the transverse parameters are expected. These parameters correspond to the width of the $x$-dependent Gaussian functions
of the various terms in Eq.~\eqref{Fpi_sum_FockComp2} and their contributions to the form factor are modulated only by the integral over $x$.

\begin{figure}[ht]
    \centering
    \includegraphics[width=0.97\textwidth]{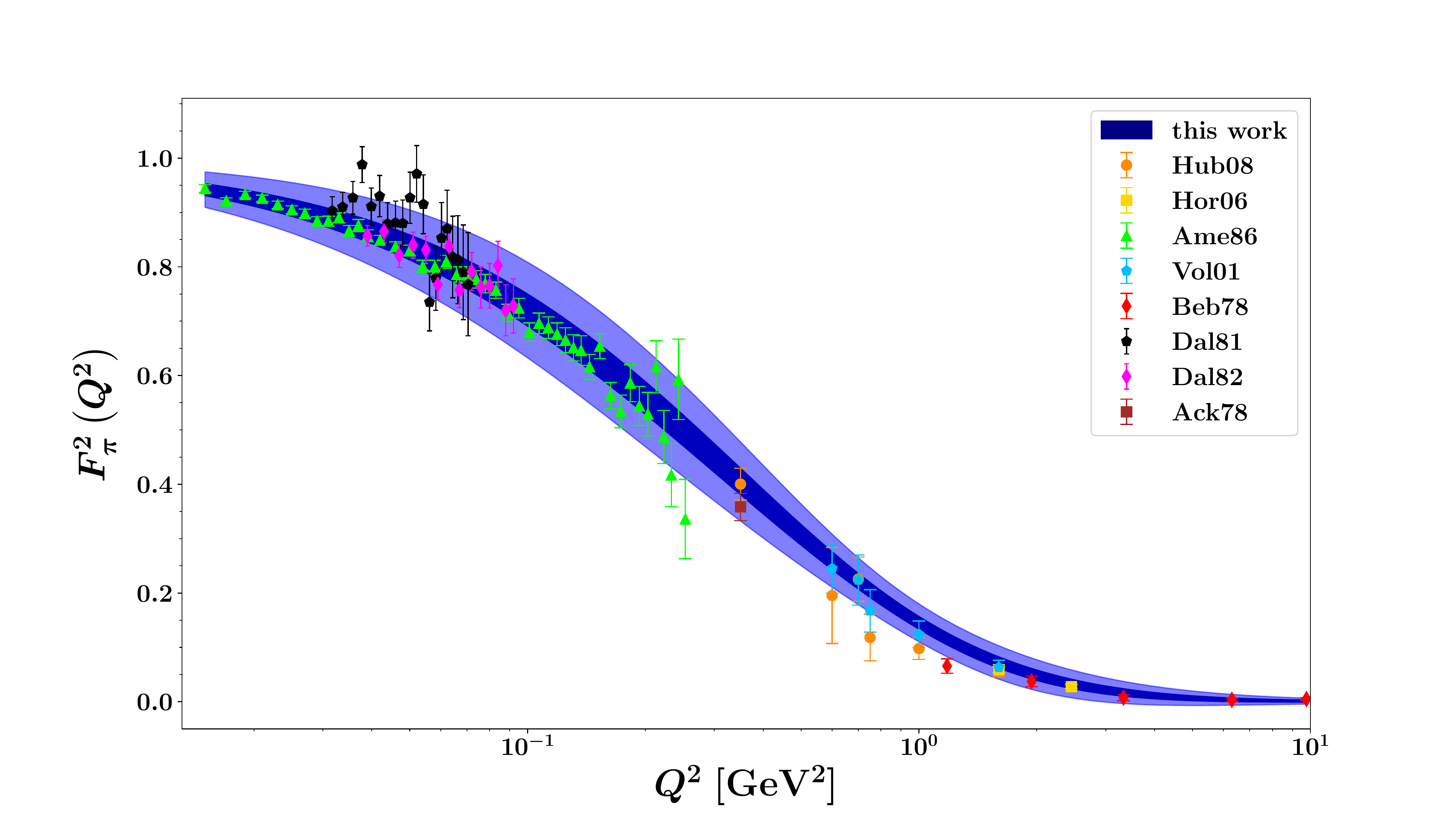}
    \caption{Fit results for the square of the pion electromagnetic form factor as function of $Q^2$.
    The dark (light) blue band shows the 68\% (99.7\%) of the replicas. The experimental data correspond to 
    Hub08~\cite{JeffersonLab:2008jve},
    Hor06~\cite{JeffersonLabFpi-2:2006ysh},
    Ame86~\cite{NA7:1986vav},
    Vol01~\cite{JeffersonLabFpi:2000nlc},
    Beb78~\cite{Bebek:1977pe},
   Dal82~\cite{Dally:1981ur,Dally:1982zk},
    Ack78~\cite{Ackermann:1977rp}.
    }
    \label{FF_results_figure}
    \end{figure}
In Fig.~\ref{FF_results_figure}, we show the results for the square of the pion e.m. form factor obtained from the fit with the bootstrap method.
The inner (dark blue) band represents the 
68\% uncertainty, while the external (light blue) band shows the 
99.7\% uncertainty.
Agreement with the different data sets is qualitatively evident.
We stress that the two bands incorporate the error propagation of the PDF parameters, representing therefore more than just the experimental uncertainty on the form factor.

\section{Conclusions}\label{Sec_Conclusions}
In this work, we presented an extraction of the pion parton distribution functions and pion electromagnetic form factor 
using  a new parametrization  for the pion light-front wave functions (LFWFs). At the initial scale of the model,
we considered 
  the $q\bar q$, $q\bar q q \bar q$, $q\bar qg$, and $q\bar q gg$ components of the pion state, 
  spanning a larger basis of states  than in existing light-front model calculations. 
  We inferred the parametrization in the longitudinal-momentum space from the pion distribution amplitudes, while we used 
  a modified  $x$-dependent gaussian Ansatz for
  the transverse-momentum dependent part. The functional form of the LFWF for each parton configuration was chosen so that the fit of the collinear PDFs does not contain any spurious dependence of the parameters in the transverse-momentum space. 
  The collinear parameters were fitted to available Drell-Yan and photon-production data within the xFitter framework, using the bootstrap method for the error analysis.  The quality of the fit of the pion PDFs is comparable to existing extractions in literature with the main differences  for the gluon and sea contributions that are less constrained by the available data.  However, comparing our results for the pion PDFs with other model calculations where the gluon and/or the sea contributions are generated only through perturbative evolution, we can  appreciate the effects of including  non-perturbative sea and gluon contributions in the light-front Fock expansion at the hadronic scale.
  Once determined the collinear parameters, we were able to fix the   
  residual  transverse-momentum dependent parameters from a fit to the available data on the pion electromagnetic form factor. 
The fit was performed with a  bootstrap method  that  incorporates 
the propagation of the uncertainties for the collinear parameters in the error band of the electromagnetic form factor and provides the  correlation 
matrix of the whole set of collinear and transverse parameters.

The procedure outlined in this work  represents a proof-of-principle for  a unified description of hadron distribution functions from inclusive and exclusive processes that involve both the longitudinal and transverse momentum motion of partons inside hadrons.
The final goal will be to include also the transverse-momentum dependent parton distributions (TMDs) and generalized parton distributions (GPDs)  in  a global fit, capitalizing on the extended database expected from upcoming experiments at JLab, COMPASS++/AMBER, and future electron-ion colliders and  going a step further than existing analysis  that so far considered only PDFs and TMDs simultaneously~\cite{Barry:2023qqh} or focused  on TMDs~\cite{Vladimirov:2019bfa,Cerutti:2022lmb} and  GPDs~\cite{Chavez:2021llq} separately.

\acknowledgments
We are grateful to V. Bertone and I. Novikov for valuable discussions and for the help in using  the xFitter framework.
We thank G. Bozzi for a careful reading of the manuscript and useful comments.
We acknowledge all the groups who provided us with their results for the 
pion PDFs at different scales, not always available in the 
original publications: in particular, D. Binosi  for 
the BSWF results, C. Bourrelly and J.-C. Peng for the BDP 
results, J.Lan for the BLFQ results, P. Barry for the JAM 
extractions, and I. Novikov for the xFitter results.
The work of B.P. and S.V. is supported
by the European Union’s Horizon 2020 programme under
Grant Agreement No. 824093 (STRONG2020).

\appendix

\section{LFWA overlap representation of the pion parton distribution function}\label{App_LFWA_PDFs}
In this appendix, we report  the overlap representation of the pion PDF in terms of the LFWAs corresponding to the parton configuration
with zero partons’ orbital angular momentum in Eqs.~\eqref{Eq_qq_State_lz=0}-\eqref{Eq_qqqq_State}. 
We can write the pion PDFs as the sum of the contributions from each parton configuration, i.e.,
\begin{align}
    f_1^{v}(x) & = f_{1, q \bar{q}}^{v}(x) + f_{1, q \bar{q} g}^{v}(x) + f_{1, q \bar{q} gg}^{v}(x) + \sum_{\left\{ \mathcal{s} \bar{\mathcal{s}}\right\}} f_{1, q \bar{q}  \left\{ \mathcal{s}\bar{\mathcal{s}}\right\}}^{v}(x),\\
    f_1^{g}(x) & = f_{1, q \bar{q} g}^{g}(x) + f_{1, q \bar{q} gg}^{g}(x), \\
    f_1^{S}(x) & = 2 \sum_{\left\{ \mathcal{s} \bar{\mathcal{s}}\right\}} f_{1, q \bar{q} \left\{\mathcal{s}\bar{\mathcal{s}}\right\}}^{S}(x), 
    \end{align}
where
\begin{align}
	 f^{v}_{1,q\bar q}(x) &= 4 \int \text{d[}1\text{]}\text{d[}2\text{]} \sqrt{x_1 x_2} \delta(x-x_1) |\psi^{(1)}_{q\bar{q}}(1,2)|^2 ,
\\
	f^{v}_{1, q \bar{q} g}(x) &= 4 \int \text{d[}1\text{]}\text{d[}2\text{]}\text{d[}3\text{]}  \sqrt{x_1 x_2 x_3} \delta(x-x_1) |\psi^{(1)}_{q\bar{q}g}(1,2,3)|^2, \\
 f^{v}_{1,q\bar{q}gg}(x) &= 16\int \text{d[}1\text{]}\text{d[}2\text{]}\text{d[}3\text{]}\text{d[}4\text{]} \sqrt{x_1 x_2x_3x_4} \delta(x-x_1) \left[|\psi^{(1)}_{q\bar{q}gg}(1,2,3,4)|^2  + |\psi^{(2)}_{q\bar{q}gg}(1,2,3,4)|^2 \right], \\
	f^{v}_{1,q\bar{q}\left\{\mathcal{s}\bar{\mathcal{s}}\right\}}(x) &= 8 \int \text{d[}1\text{]}\text{d[}2\text{]}\text{d[}3\text{]}\text{d[}4\text{]} \sqrt{x_1 x_2x_3x_4} \delta(x-x_1) \left[|\psi^{(1)}_{q\bar{q}\mathcal{s}\bar{\mathcal{s}}}(1,2,3,4)|^2  + |\psi^{(2)}_{q\bar{q}\mathcal{s}\bar{\mathcal{s}}}(1,2,3,4)|^2 + \frac{1}{2} |\psi^{(3)}_{q\bar{q}\mathcal{s}\bar{\mathcal{s}}}(1,2,3,4)|^2\right],\nonumber\\
&\\
f_{1, q \bar{q}g}^{g}(x)&= 2 \int \text{d[}1\text{]}\text{d[}2\text{]}\text{d[}3\text{]} \sqrt{x_1x_2x_3}\delta(x-x_3) |\psi^{(1)}_{q\bar{q}g}(1,2,3)\big|^2 , \\
	f^{g}_{1,q\bar{q}gg} (x)&= 16 \int \text{d[}1\text{]}\text{d[}2\text{]}\text{d[}3\text{]}\text{d[}4\text{]} \sqrt{x_1 x_2x_3x_4} \delta(x-x_3) \left[|\psi^{(1)}_{q\bar{q}gg}(1,2,3,4)|^2  + |\psi^{(2)}_{q\bar{q}gg}(1,2,3,4)|^2 \right], \\
	f^{S}_{1,q\bar{q}\left\{\mathcal{s}\bar{\mathcal{s}}\right\}}(x) &= 4 \int \text{d[}1\text{]}\text{d[}2\text{]}\text{d[}3\text{]}\text{d[}4\text{]} \sqrt{x_1 x_2x_3x_4} \delta(x-x_3) \left[|\psi^{(1)}_{q\bar{q}\mathcal{s}\bar{\mathcal{s}}}(1,2,3,4)|^2  + |\psi^{(2)}_{q\bar{q}\mathcal{s}\bar{\mathcal{s}}}(1,2,3,4)|^2 + \frac{1}{2} |\psi^{(3)}_{q\bar{q}\mathcal{s}\bar{\mathcal{s}}}(1,2,3,4)|^2\right].\nonumber\\
&
\end{align}

\section{LFWA overlap representation of the pion form factor}\label{App_LFWA_FFs}

In this appendix, we report  the overlap representation of the pion form factor in terms of the LFWAs corresponding to the parton configuration
with zero partons’ orbital angular momentum in Eqs.~\eqref{Eq_qq_State_lz=0}-\eqref{Eq_qqqq_State}. 
The contributions from each parton configuration in Eq.~\eqref{Fpi_sum_FockComp} read

\begin{align}
	 F_{\pi ,q\bar{q}}(Q^2) &= 2 \int \text{d[}1\text{]}\text{d[}2\text{]} \sqrt{x_1 x_2}\psi^{*(1)}_{q\bar{q}}(x_1,x_2, \boldsymbol{k}_{\perp 1} + (1 - x_1)\boldsymbol{q}_{\perp}, \boldsymbol{k}_{\perp 2} - x_2 \boldsymbol{q}_{\perp})\psi^{(1)}_{q\bar{q}}(1,2) ,
\\
	 F_{\pi ,q\bar{q} g}(Q^2) &= 2 \int \text{d[}1\text{]}\text{d[}2\text{]}\text{d[}3\text{]}  \sqrt{x_1 x_2 x_3} \psi^{*(1)}_{q\bar{q}g}(x_1,x_2,x_3,\boldsymbol{k}_{\perp 1} +(1-x_1)\boldsymbol{q}_{\perp}, \boldsymbol{k}_{\perp 2} -x_2\boldsymbol{q}_{\perp}, \boldsymbol{k}_{\perp 3} -x_3\boldsymbol{q}_{\perp}) \psi^{(1)}_{q\bar{q}g}(1,2,3),\\ F_{\pi ,q\bar{q} gg}(Q^2) &= 4 \int \text{d[}1\text{]}\text{d[}2\text{]}\text{d[}3\text{]}\text{d[}4\text{]} \sqrt{x_1 x_2 x_3 x_4} \nonumber\\
    \nonumber &\times\biggl[\psi^{*(1)}_{q\bar{q} gg}(x_1,x_2, x_3, x_4, \boldsymbol{k}_{\perp 1} + (1 - x_1)\boldsymbol{q}_{\perp}, \boldsymbol{k}_{\perp 2} - x_2 \boldsymbol{q}_{\perp}, \boldsymbol{k}_{\perp 3} - x_3 \boldsymbol{q}_{\perp},  \boldsymbol{k}_{\perp 4} - x_4 \boldsymbol{q}_{\perp})\psi^{(1)}_{q\bar{q} gg}(1,2,3,4) \\
    \nonumber &+\psi^{*(1)}_{q\bar{q} gg}(x_1,x_2, x_4, x_3, \boldsymbol{k}_{\perp 1} + (1 - x_1)\boldsymbol{q}_{\perp}, \boldsymbol{k}_{\perp 2} - x_2 \boldsymbol{q}_{\perp}, \boldsymbol{k}_{\perp 4} - x_4 \boldsymbol{q}_{\perp},  \boldsymbol{k}_{\perp 3} - x_3 \boldsymbol{q}_{\perp})\psi^{(1)}_{q\bar{q} gg}(1,2,3,4) \\
    \nonumber &+\psi^{*(2)}_{q\bar{q} gg}(x_1,x_2, x_3, x_4, \boldsymbol{k}_{\perp 1} + (1 - x_1)\boldsymbol{q}_{\perp}, \boldsymbol{k}_{\perp 2} - x_2 \boldsymbol{q}_{\perp}, \boldsymbol{k}_{\perp 3} - x_3 \boldsymbol{q}_{\perp},  \boldsymbol{k}_{\perp 4} - x_4 \boldsymbol{q}_{\perp})\psi^{(2)}_{q\bar{q} gg}(1,2,3,4) \\
    &-\psi^{*(2)}_{q\bar{q} gg}(x_1,x_2, x_4, x_3, \boldsymbol{k}_{\perp 1} + (1 - x_1)\boldsymbol{q}_{\perp}, \boldsymbol{k}_{\perp 2} - x_2 \boldsymbol{q}_{\perp}, \boldsymbol{k}_{\perp 4} - x_4 \boldsymbol{q}_{\perp},  \boldsymbol{k}_{\perp 3} - x_3 \boldsymbol{q}_{\perp})\psi^{(2)}_{q\bar{q} gg}(1,2,3,4) \biggr] ,\\
	 \nonumber F_{\pi ,q\bar{q} \left\{\mathcal{s}\bar{\mathcal{s}}\right\}}(Q^2) &= 4 \int \text{d[}1\text{]}\text{d[}2\text{]}\text{d[}3\text{]}\text{d[}4\text{]} \sqrt{x_1 x_2 x_3 x_4} \\
    \nonumber &\times\biggl[\psi^{*(1)}_{q\bar{q} \mathcal{s} \bar{\mathcal{s}}}(x_1,x_2, x_3, x_4, \boldsymbol{k}_{\perp 1} + (1 - x_1)\boldsymbol{q}_{\perp}, \boldsymbol{k}_{\perp 2} - x_2\boldsymbol{q}_{\perp}, \boldsymbol{k}_{\perp 3} - x_3 \boldsymbol{q}_{\perp},  \boldsymbol{k}_{\perp 4} - x_4 \boldsymbol{q}_{\perp})\psi^{(1)}_{q\bar{q} \mathcal{s} \bar{\mathcal{s}}}(1,2,3,4) \\
    \nonumber &+\psi^{*(2)}_{q\bar{q} \mathcal{s} \bar{\mathcal{s}}}(x_1,x_2, x_3, x_4, \boldsymbol{k}_{\perp 1} + (1 - x_1)\boldsymbol{q}_{\perp}, \boldsymbol{k}_{\perp 2} - x_2\boldsymbol{q}_{\perp}, \boldsymbol{k}_{\perp 3} - x_3 \boldsymbol{q}_{\perp},  \boldsymbol{k}_{\perp 4} - x_4 \boldsymbol{q}_{\perp})\psi^{(2)}_{q\bar{q} \mathcal{s} \bar{\mathcal{s}}}(1,2,3,4) \\
    &+ \frac{1}{2}\psi^{*(3)}_{q\bar{q} \mathcal{s} \bar{\mathcal{s}}}(x_1,x_2, x_3, x_4, \boldsymbol{k}_{\perp 1} + (1 - x_1)\boldsymbol{q}_{\perp}, \boldsymbol{k}_{\perp 2} - x_2\boldsymbol{q}_{\perp}, \boldsymbol{k}_{\perp 3} - x_3 \boldsymbol{q}_{\perp},  \boldsymbol{k}_{\perp 4} - x_4 \boldsymbol{q}_{\perp})\psi^{(3)}_{q\bar{q} \mathcal{s} \bar{\mathcal{s}}}(1,2,3,4) \biggr].
\end{align}
%

\end{document}